\documentclass{aa}  

\usepackage{graphicx}
\usepackage{gensymb}
\usepackage{txfonts}
\usepackage{scrextend}
\usepackage{amssymb}
\usepackage{mathrsfs}
\usepackage[dvipsnames]{xcolor}

\usepackage{array}
\newcolumntype{T}{>{\tiny}c} % define a new column type for \tiny

\providecommand{\bjdtdb}{\ensuremath{\rm {BJD_{TDB}}}}
\providecommand{\feh}{\ensuremath{\left[{\rm Fe}/{\rm H}\right]}}
\providecommand{\teff}{\ensuremath{T_{\rm eff}}}

\providecommand{\mj}{\ensuremath{\,M_{\rm J}}}
\providecommand{\rj}{\ensuremath{\,R_{\rm J}}}

\newcommand{\logg}{\mbox{$\log g_*$}\xspace}
\newcommand{\vsini}{\mbox{$v \sin i_{*}$}\xspace}
\newcommand{\kms}{\mbox{km\,s$^{-1}$}\xspace}
\newcommand{\ms}{\mbox{m\,s$^{-1}$}\xspace}

\newcommand{\mearth}{\mbox{$\mathrm{M_{\oplus}}$}\xspace}
\newcommand{\mjup}{\mbox{$\mathrm{M_{\rm Jup}}$}\xspace}
\newcommand{\rjup}{\mbox{$\mathrm{R_{\rm Jup}}$}\xspace}

\newcommand{\mstar}{\mbox{$M_{*}$}\xspace}
\newcommand{\rstar}{\mbox{$R_{*}$}\xspace}

\newcommand{\msol}{\mbox{$\mathrm{M_\odot}$}\xspace}
\newcommand{\rsol}{\mbox{$\mathrm{R_\odot}$}\xspace}

\newcommand{\lsol}{\mbox{$\mathrm{L_\odot}$}\xspace}

\begin{document} 

\title{NGTS-13b: A hot 4.8 Jupiter-mass planet transiting a subgiant star}

\titlerunning{NGTS-13b: A hot massive Jupiter}
\authorrunning{N. Grieves, et al.}
   \author{Nolan Grieves
   \inst{\ref{inst:geneva}}\fnmsep\thanks{\email{nolan.grieves@unige.ch}}
    \and Louise D. Nielsen \inst{\ref{inst:geneva}}
    \and Jose I. Vines \inst{\ref{inst:chil}}
    \and Edward M. Bryant\inst{\ref{inst:war},\ref{inst:warexo}}
    \and Samuel Gill\inst{\ref{inst:war},\ref{inst:warexo}}
    \and Fran\c{c}ois Bouchy \inst{\ref{inst:geneva}}
    \and Monika Lendl\inst{\ref{inst:geneva}}
    \and Daniel Bayliss\inst{\ref{inst:war},\ref{inst:warexo}}
    \and Philipp Eigmueller\inst{\ref{inst:gac}}
    \and  Damien Segransan \inst{\ref{inst:geneva}}
    % Alphabetic:
    \and Jack S. Acton \inst{\ref{inst:leic}}
    \and David~R.~Anderson\inst{\ref{inst:war},\ref{inst:warexo}}
    \and Matthew R. Burleigh \inst{\ref{inst:leic}}
    \and Sarah L. Casewell \inst{\ref{inst:leic}}
    \and Alexander Chaushev \inst{\ref{inst:tu}}
    \and Benjamin F. Cooke\inst{\ref{inst:war},\ref{inst:warexo}}
    \and Edward Gillen  \inst{\ref{inst:qmul},\ref{inst:camb}}\thanks{Winton Fellow}
    \and Michael R. Goad \inst{\ref{inst:leic}}
    \and Maximilian N. G{\"u}nther\inst{\ref{inst:mit}}\thanks{Juan Carlos Torres Fellow}
    \and Beth A. Henderson \inst{\ref{inst:leic}}
    \and Aleisha Hogan \inst{\ref{inst:leic}}
    \and James S. Jenkins\inst{\ref{inst:chil},\ref{inst:cata}}
    \and Douglas R. Alves \inst{\ref{inst:chil}}
    \and Andr\'es Jord\'an\inst{\ref{inst:adolfo},\ref{inst:mill}}
    \and James McCormac \inst{\ref{inst:war},\ref{inst:warexo}}
    \and Maximiliano Moyano \inst{\ref{inst:antchil}}
    \and Didier Queloz \inst{\ref{inst:camb}}
    \and Liam Raynard \inst{\ref{inst:leic}}
    \and Julia V. Seidel \inst{\ref{inst:geneva}}
    \and Alexis M. S. Smith \inst{\ref{inst:gac}}
    \and Rosanna H. Tilbrook \inst{\ref{inst:leic}}
    \and Stephane Udry \inst{\ref{inst:geneva}}
    \and Richard G.\ West \inst{\ref{inst:war},\ref{inst:warexo}}
    \and Peter J.\ Wheatley \inst{\ref{inst:war},\ref{inst:warexo}}
    }
      
 \institute{%1
   Observatoire de Gen{\`e}ve, Universit{\'e} de Gen{\`e}ve, Chemin Pegasi 51b, 1290 Sauverny, Switzerland \label{inst:geneva}
         \and %2
         Departamento de Astronom\'ia, Universidad de Chile, Casilla 36-D, Santiago, Chile \label{inst:chil}
         \and        
         Department of Physics, University of Warwick, Gibbet Hill Road, Coventry, CV4 7AL, UK \label{inst:war}
         \and 
         Centre for Exoplanets and Habitability, University of Warwick, Gibbet Hill Road, Coventry, CV4 7AL, UK \label{inst:warexo}
         \and 
         Institute of Planetary Research, German Aerospace Center, Rutherfordstrasse 2, 12489 Berlin, Germany \label{inst:gac}
         \and
         School of Physics and Astronomy, University of Leicester, LE1 7RH, UK \label{inst:leic}
         \and
         Center for Astronomy and Astrophysics, TU Berlin, Hardenbergstr. 36, D-10623 Berlin, Germany \label{inst:tu}
         \and
         Astronomy Unit, Queen Mary University of London, Mile End Road, London E1 4NS, UK  \label{inst:qmul}
         \and 
         Astrophysics Group, Cavendish Laboratory, J.J. Thomson Avenue, Cambridge CB3 0HE, UK \label{inst:camb}
         \and
         Department of Physics, Kavli Institute for Astrophysics and Space Research, MIT, 77 Mass. Ave, Cambridge, MA 02139, USA \label{inst:mit}
         \and
         Centro de Astrof\'isica y Tecnolog\'ias Afines (CATA), Casilla 36-D, Santiago, Chile \label{inst:cata}
         \and 
         Facultad de Ingenier\'ia y Ciencias, Universidad Adolfo Ib\'a\~{n}ez, Av.Diagonal las Torres 2640, Pe\~{n}alol\'en, Santiago, Chile \label{inst:adolfo}
         \and
         Millennium Institute for Astrophysics, Chile \label{inst:mill}
         \and 
         Instituto de Astronom\'ia, Universidad Cat\'olica del Norte, Angamos 0610, 1270709, Antofagasta, Chile \label{inst:antchil}
         }
   \date{Received 2 October 2020; accepted 6 January 2020}

% \abstract{}{}{}{}{} 
% 5 {} token are mandatory
 
      \abstract{We report the discovery of the massive hot Jupiter NGTS-13b by the Next Generation Transit Survey (NGTS). The V = 12.7 host star is likely in the subgiant evolutionary phase with \logg = 4.04\,$\pm$\,0.05, $T_{\rm eff}$ = 5819\,$\pm$\,73 \,K, $M_*$ = 1.30$^{+0.11}_{-0.18}$\,\msol, and $R_*$ = 1.79\,$\pm$\,0.06\,\rsol. NGTS detected a transiting planet with a period of $P$ = 4.12 days around the star, which was later validated with the Transiting Exoplanet Survey Satellite (TESS; TIC 454069765). We confirm the planet using radial velocities from the CORALIE spectrograph. Using NGTS and TESS full-frame image photometry combined with CORALIE radial velocities we determine NGTS-13b to have a radius of $R_P$ = 1.142\,$\pm$\,0.046\,\rjup, mass of $M_P$ = 4.84\,$\pm$\,0.44\,\mjup, and eccentricity $e$ = 0.086\,$\pm$\,0.034. Some previous studies suggest that $\sim$4\,\mjup may be a border between two separate formation scenarios (e.g., core accretion and disk instability) and that massive giant planets share similar formation mechanisms as lower-mass brown dwarfs. NGTS-13b is just above 4\,\mjup making it an important addition to the statistical sample needed to understand the differences between various classes of substellar companions. The high metallicity, $\feh$ = 0.25\,$\pm$\,0.17, of NGTS-13 does not support previous suggestions that massive giants are found preferentially around lower metallicity host stars, but NGTS-13b does support findings that more massive and evolved hosts may have a higher occurrence of close-in massive planets than lower-mass unevolved stars.}



   \maketitle
%
%-------------------------------------------------------------------

\section{Introduction}\label{sec:intro}

The first extrasolar planets discovered were a new class of planets not found in our solar system: hot Jupiters -- Jupiter-mass planets with orbital periods $\leq$ 10 days, e.g., \citet{MayorQueloz1995}. Planet formation and evolution theories were abruptly altered to account for these distinct systems, e.g., \citet{Boss1996}. However, there is still no consensus on the predominant formation mechanism of these massive close-in planets. Currently three main hypotheses are suggested as possible mechanisms for how hot Jupiters could have formed and now reside so close to their host star: in situ formation, disk migration, and high-eccentricity tidal migration. For a review of these scenarios see \citet{DawsonJohnson2018}.

There are two mechanisms proposed for giant planet formation in general: gravitational instability -- through its own self-gravity the protoplanetary disk fragments into giant gaseous protoplanets that contract and collapse to form giant planets \citep[e.g.,][]{Boss1997}, and core accretion -- planetesimal collisions form rocky protoplanet cores that accrete gas from the protoplanetary disk \citep[e.g.,][]{Pollack1996}. In situ gravitational instability is not plausible at short orbital periods because the disk's fast rotation and high temperature (thermal pressure) support the local gas against gravitational collapse \citep{DawsonJohnson2018}. Core accretion can only occur if there is enough mass in the feeding zone to grow a sufficiently large core \citep[$\sim$10 $\mearth$;][]{Rafikov2006,Lee2014,Piso2015} within the disk's lifetime, but at short orbital periods the feeding zone is smaller (smaller Hill radii) making it implausible to form a massive core in normal disk conditions \citep{LeeChiang2016}. However, in situ core accretion of hot Jupiters may be possible with highly efficient radial drift of solids to the inner disk \citep{DawsonJohnson2018}.

A more common line of thought is that hot Jupiters formed at larger orbital periods and migrated inward. One mechanism for this migration is via disk migration by exchanging angular momentum with the disk through corotation and Lindblad torques \citep[e.g.,][]{Lin1996}. Another proposed mechanism for migration is high-eccentricity tidal migration, where a giant planet has orbital angular momentum extracted from it by a perturber, pushing the planet to a highly elliptical orbit, then the planet reduces its orbital energy through tidal interactions with the central star causing it to circularize into a close-in orbit \citep[e.g.,][]{RasioFord1996}.

These various formation theories should have differing effects on the characteristics of hot Jupiters. Slow and smooth migration in the circumstellar disk from simple angular momentum loss should produce aligned and circularized orbits \citep{DAngelo2003}, but more dynamic migration, such as high-eccentricity tidal migration, should produce planets with higher eccentricites in more misaligned orbits \citep[e.g.,][]{Triaud2010}. However, for hot Jupiters with orbital periods shorter than 5 days the eccentricity will be erased in a few billion years due to tidal circularization \citep{Adams2006}. Orbital misalignment, a difference in a star's rotation axis and a planet’s orbital axis (obliquity), evolves more slowly than eccentricity \citep{Hut1981,Winn2005,BarkerOgilvie2009} and therefore can be detected in relatively older stars. There are various methods for measuring obliquity, but particularly for transiting planets radial velocity (RV) observations can be taken during the transit to find the spin-orbit alignment via the Rossiter-McLaughin effect \citep{Rossiter1924,McLaughlin1924}. 

The parameter distribution of the entire sample of hot Jupiters such as their semi-major axes and ages also give insight into which formation mechanisms dominate in general and in specific environments. It is thus important to continue to build a sample of planets (and host stars) with well-characterized parameters. When a planet is detected by both the transit and RV methods multiple characteristics of the planet can be well-constrained compared to only detecting it through one method. Thus, large photometric surveys are valuable as they can view large samples of stars and identify those with transiting planets, which can then be followed up with RV instruments that typically can only observe one star at a time. 

The  Next  Generation  Transit  Survey  \citep[NGTS;][]{Wheatley2018} is a ground-based photometric survey located in Chile at the European Southern Observatory (ESO) Paranal site. NGTS consists of twelve fully robotic telescopes each with a 20\,cm photometric aperture and a wide 8 deg$^{2}$ field-of-view making the survey capable of observing a large sample of stars. Here we present the discovery of a massive hot Jupiter orbiting a subgiant star. This planet was first detected by NGTS, subsequently verified by the space-based all-sky Transiting Exoplanet Survey Satellite \citep[TESS;][]{Ricker2015}, and confirmed through follow-up RV measurements with the CORALIE \citep{Queloz2001} spectrograph. In Section~\ref{sec:obs} we present our observations, in Section~\ref{sec:analysis} we describe our analysis and present our results, in Section~\ref{sec:discussion} we discuss our results, and we present our conclusions in Section~\ref{sec:conc}.

%--------------------------------------------------------------------
\section{Observations}\label{sec:obs}

\begin{figure}
  \centering
  \includegraphics[width=0.4\textwidth]{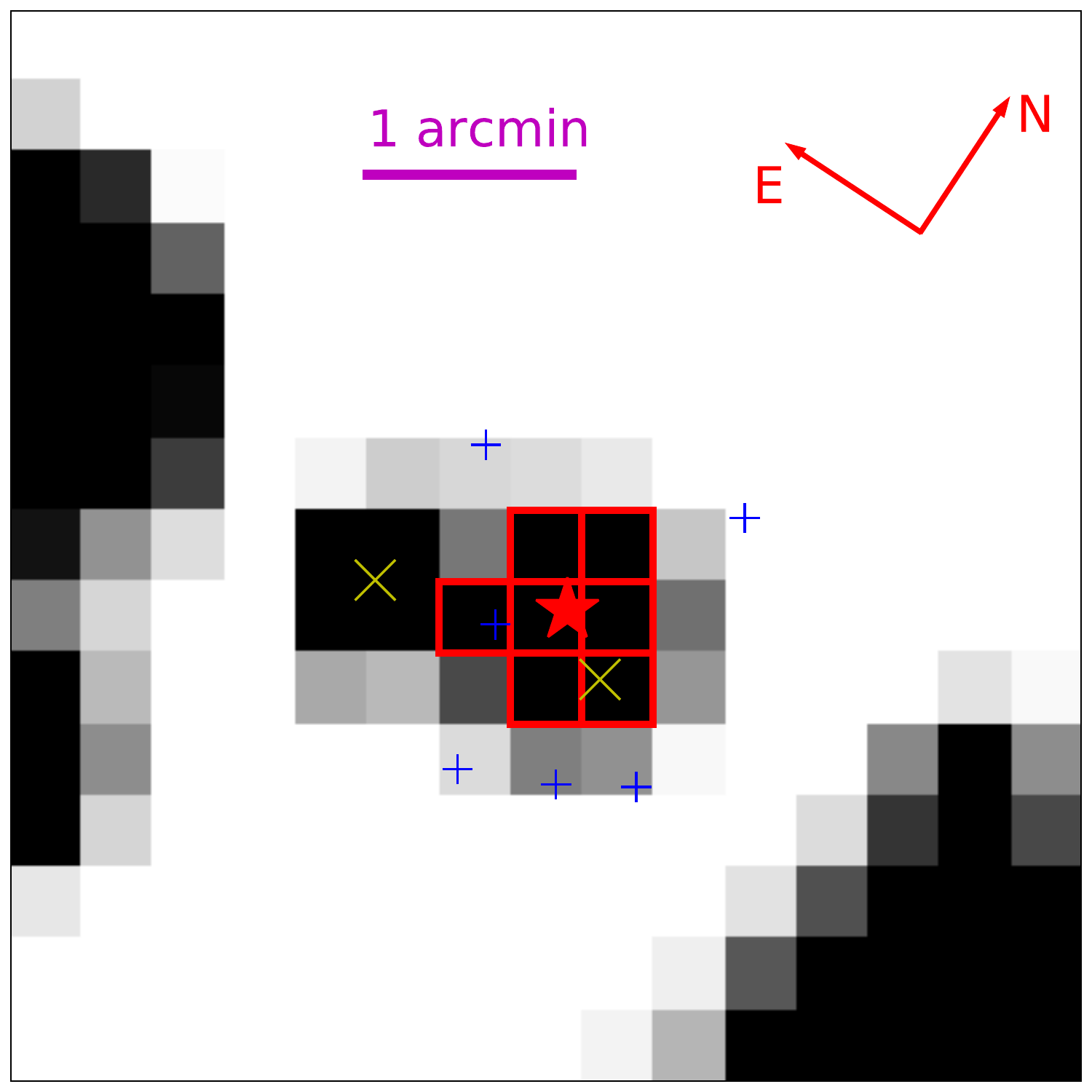}
  \caption{TESS Full-Frame Image cutout (15 $\times$ 15 pixels) in the region of NGTS-13 (TIC 454069765; T=12.11) from which the TESS light curve was generated. The red boxes indicate the target pixel aperture mask used to extract the photometry and the white boxes without a border are the pixels used to determine the sky background. The red star indicates the position of NGTS-13. The yellow crosses indicate TIC 454069767 ($\Delta$T=1.68, 22.5” west) and TIC 454069772 ($\Delta$T=0.6, 57” southeast). Blue pluses denote the positions of stars in $GAIA$ DR2 within 1’ of NGTS-13.}
  \label{fig:tesscut}
\end{figure}

\begin{figure*}
  \centering
  \includegraphics[width=0.9\textwidth]{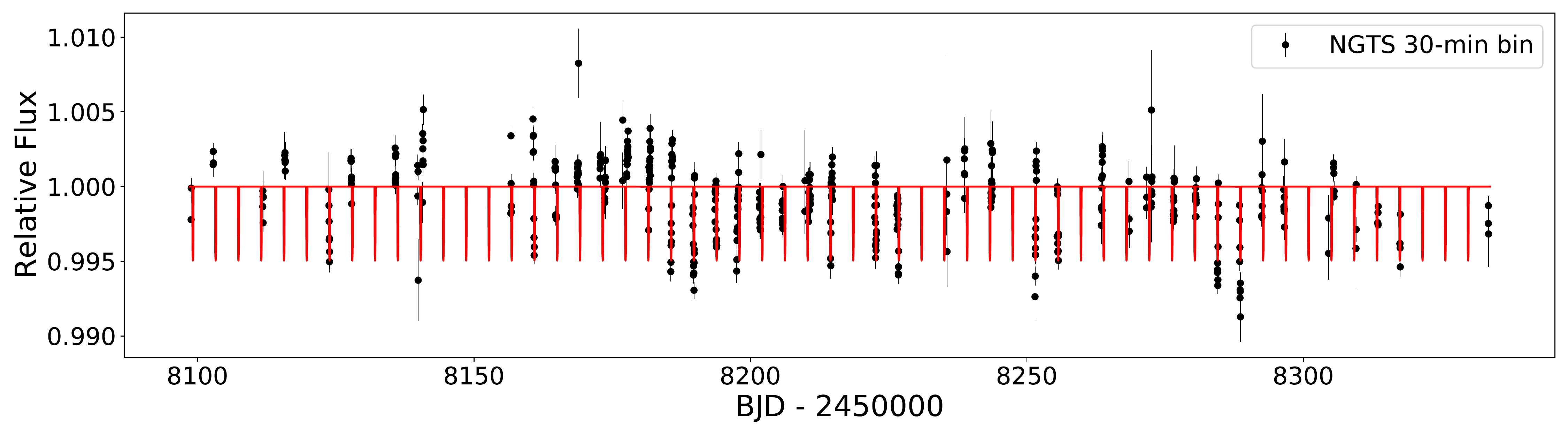}
  \includegraphics[width=0.9\textwidth]{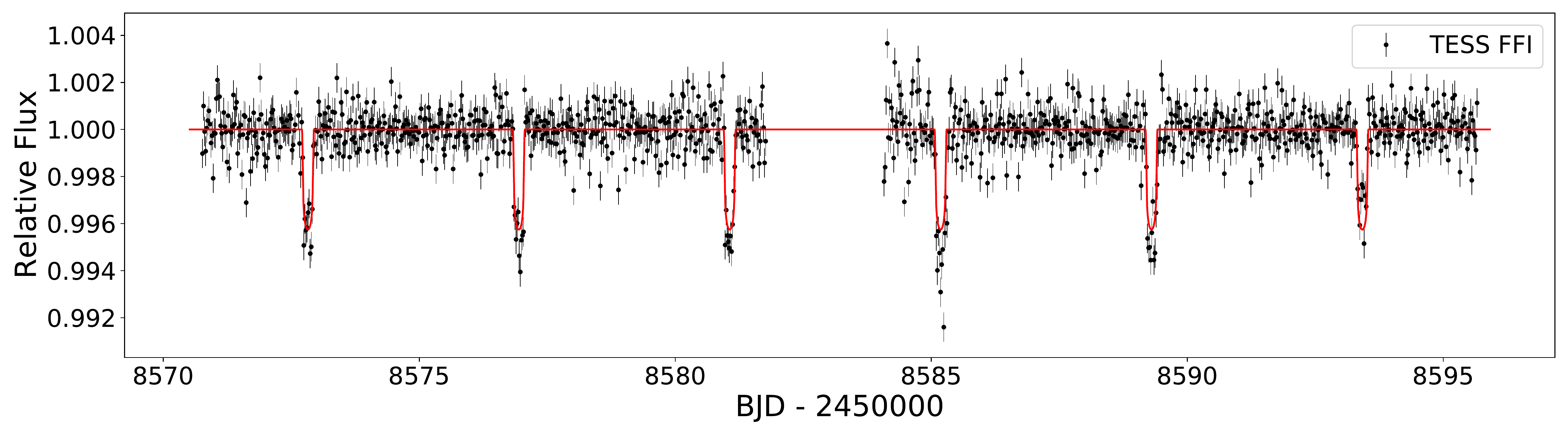}
  \includegraphics[width=0.99\textwidth]{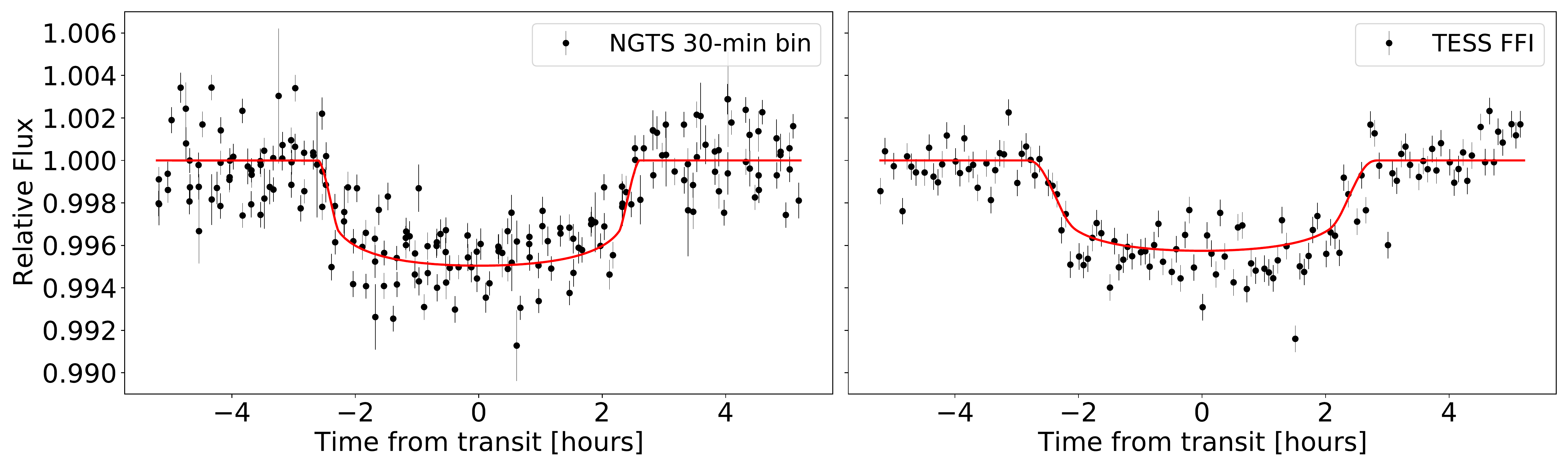}
  \caption{\textit{Top}: NGTS observations (binned to 30-minutes for visual purposes) with the EXOFASTv2 model (see Section~\ref{sec:glob}) in red. \textit{Middle}: TESS Full Frame Image (FFI) observations after the detrending described in Section \ref{sec:tessobs} was applied with the the EXOFASTv2 model in red. \textit{Bottom}: NGTS binned to 30-minutes and TESS FFI photometry phased to the 4.12 day period of NGTS-13b. The NGTS data are binned to 30-minutes and then phased to the period. The red lines show the EXOFASTv2 model.}
  \label{fig:phasetran}
\end{figure*}

\subsection{NGTS Photometry}

NGTS-13 was observed using a single NGTS telescope from UT 2017 December 10 to UT 2018 August 7. During this period 255,709 observations were taken with exposure times of 10 seconds. Image reduction and photometry were performed with the NGTS pipeline \citep{Wheatley2018} using CASUtools software\footnote{http://casu.ast.cam.ac.uk/surveys-projects/software-release}. The lightcurve was detrended and systematic effects were removed using the SysRem algorithm \citep{Tamuz2005}. 

The NGTS lightcurves were searched for periodic transit-like signals using an implementation of the box-fitting least-squares (BLS) algorithm \citep{Kovacs2002}. We detected a strong peak in the BLS periodogram at 4.12 days and folding the NGTS photometry on this period revealed a clear transit signal. 

We binned the NGTS data to 2-minute bins for our analysis as described in Section~\ref{sec:glob}. NGTS-13 has some nearby companions (see Figure~\ref{fig:tesscut}) and the NGTS photometry is expected to suffer from some dilution, which we discuss in Section~\ref{sec:dil}. The NGTS photometry binned to 30 minutes for display purposes is shown in Figure~\ref{fig:phasetran}. A full table of reduced NGTS photometric observations is available in a machine-readable format from the online journal.

\subsection{TESS Photometry} \label{sec:tessobs}
TESS observed NGTS-13 (TIC 454069765) in Sector 10 from UT 2019 March 28 to UT 2019 April 21 with 30-minute cadence Full Frame Images (FFIs). We created a lightcurve via aperture photometry of the TESS 30-minute cadence FFIs following the method set out in \citet{Gill2020}. We determined a threshold for target pixels and for background pixels based on an iterative sigma-clipping to determine the median and standard deviation of the background pixels. To exclude the neighboring T = 12.8 star (TIC 454069772, 57'' southeast, see Figure~\ref{fig:tesscut}), we only include pixels in our final aperture for which neighbouring pixels closer to the center of our target show a higher illumination. For the final pixel mask of NGTS-13, 7 pixels were selected (see Figure~\ref{fig:tesscut}). To remove the systematic trends from the TESS light curve, we mask out the transits, and then fit a linear spline interpolating across the positions of the transits. Before removing systematic trends, we estimate a signal-to-noise ratio (SNR) as the transit depth relative to the standard deviation of out-of-transit observations and find a single TESS transit SNR$\sim$5.4, which corresponds to SNR$\sim$13.3 when considering all 6 transits in the TESS data (SNR\,$\propto$\,$\sqrt{N_{\mathrm{T}}}$). The TESS photometry is displayed in Figure~\ref{fig:phasetran}. A full table of reduced TESS photometric observations is available in a machine-readable format from the online journal.

\subsection{CORALIE Spectroscopy}
\begin{figure}
  \centering
  \includegraphics[width=0.99\columnwidth]{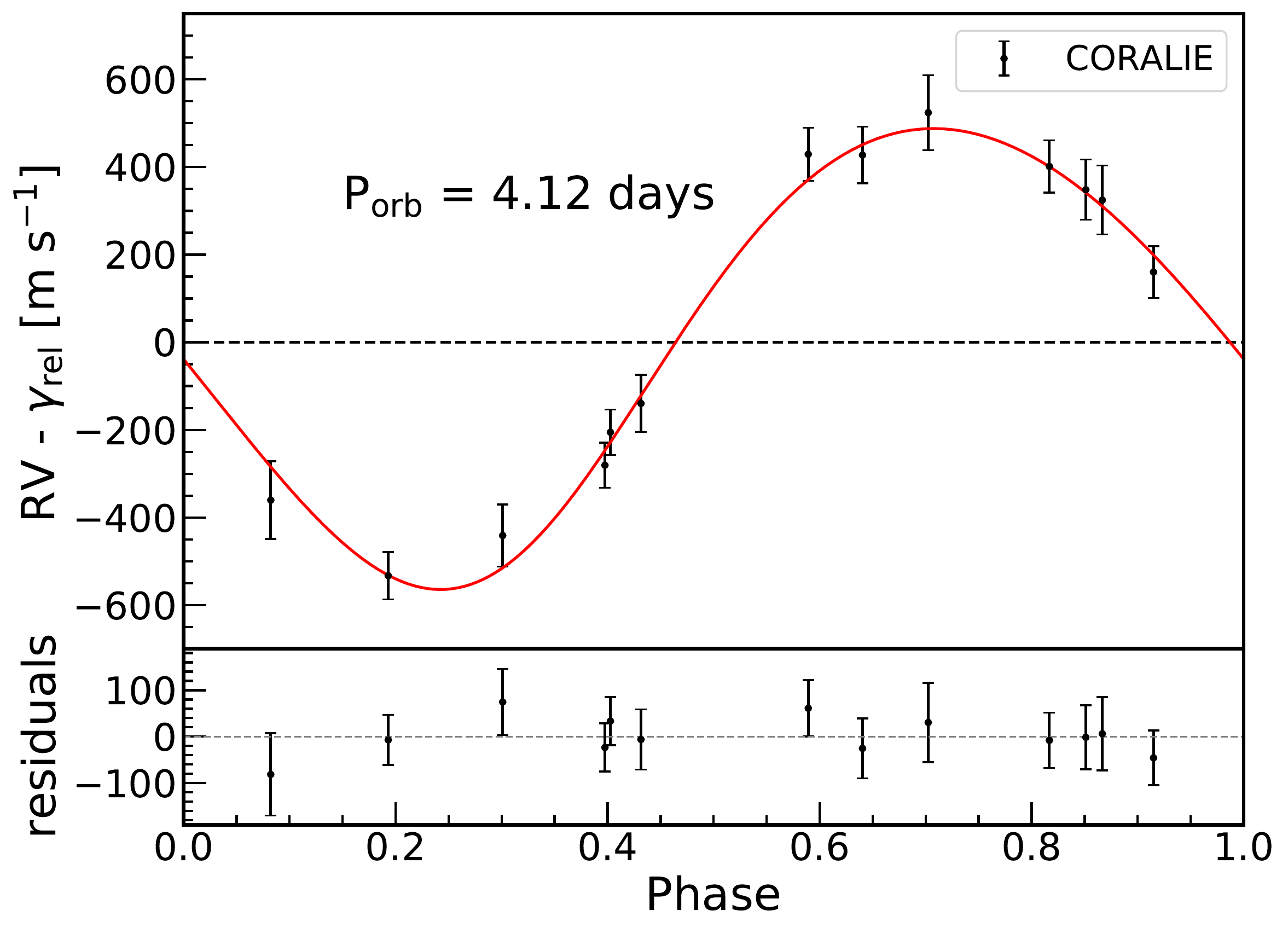}
  \caption{CORALIE Radial Velocities for NGTS-13 phased to the 4.12 day period of NGTS-13b subtracted by the relative RV offset $\gamma_{\rm rel}$ (Table~\ref{tab:global_param}). The red line shows the Keplerian fit for the planet from EXOFASTv2 (see Section~\ref{sec:glob}).}
  \label{fig:rvs}
\end{figure}

We obtained 13 observations of NGTS-13 with the high resolution echelle spectrograph CORALIE on the Swiss 1.2\,m Euler telescope \citep{Queloz2001} at La Silla Observatory, Chile between UT 2019 December 12 and UT 2020 January 23. CORALIE has a resolution of $R$\,$\sim$\,60,000 and is fed by two fibers: a 2 arcsec on-sky science fiber encompassing the star and another fiber that can either connect to a Fabry-P\'erot etalon for simultaneous wavelength calibration or on-sky for background subtraction of sky flux. We observed NGTS-13 in the simultaneous Fabry-P\'erot wavelength calibration mode using an exposure time of 2700 seconds. The spectra were reduced with the CORALIE standard reduction pipeline and RVs were computed for each epoch by cross-correlating with a binary G2 mask \citep{Pepe2002}. We co-added the 13 CORALIE spectroscopic observations by correcting each epoch to the stellar rest frame with its corresponding barycentric RV to create a single high signal-to-noise spectrum. We list the CORALIE RVs in Table~\ref{tab:coralie_rvs}. The RVs phased to the 4.12 day period are displayed in Figure~\ref{fig:rvs}.

%--------------------------------------------------------------------

\section{Analysis}\label{sec:analysis}

\begin{table}
    \centering
    \begin{tabular}{ccc}
        \hline\hline
        BJD TDB & RV [$\ms$] & RV error [$\ms$] \\
        \hline
2458829.841574 & 26749.44 & 51.75 \\
2458839.806533 & 27431.37 & 59.75 \\
2458841.801379 & 26588.94 & 71.16 \\
2458855.811893 & 27554.10 & 85.54 \\
2458857.832753 & 26497.28 & 54.13 \\
2458858.814847 & 26890.46 & 65.16 \\
2458860.806340 & 27190.32 & 59.13 \\
2458862.815160 & 26824.62 & 51.75 \\
2458863.794625 & 27457.32 & 64.68 \\
2458864.726425 & 27354.61 & 78.83 \\
2458868.781343 & 27378.28 & 68.78 \\
2458869.733432 & 26669.54 & 88.67 \\
2458871.821928 & 27459.18 & 60.68 \\
        \hline
    \end{tabular}
    \caption{CORALIE Radial Velocities for NGTS-13. Time is in Barycentric Julian Date / Barycentric Dynamical Time.}
    \label{tab:coralie_rvs}
\end{table}

\begin{table} 
    \centering
    \begin{tabular}{cccc}
        \hline\hline
        $\teff$ & $\feh$ & $\logg$  & Source \\
          K & dex & dex &   \\
        \hline
        5800\,$\pm$\,80 & 0.20\,$\pm$\,0.08 & 4.1\,$\pm$\,0.2 & Wavelet \\
        5714\,$\pm$\,110 & 0.43\,$\pm$\,0.09 & 4.11\,$\pm$\,0.12 & SpecMatch \\
        \hline
        5757\,$\pm$\,153 & 0.315\,$\pm$\,0.205 & 4.105\,$\pm$\,0.205 & Combined \\
        \hline
    \end{tabular}
    \caption{Table of stellar atmospheric parameters derived from CORALIE spectra and combining the analysis results of both SpecMatch-Emp \citep{Yee2017} and the wavelet analysis described in \citet{Gill2018,Gill2019}. We use the combined results as input priors into our global analysis.}
    \label{tab:spectra_param}
\end{table}

\begin{table*}
        \centering
        \begin{tabular}{cccc}
        \hline\hline
        Parameter & Description (Units) & Value & Source \\
        \hline
        Astrometric Properties & & & \\
        $\alpha$ & Right Ascension  & 11:44:57.66 & $Gaia$ DR2 \\ %\text{(hh:mm:ss)}
        $\delta$ & Declination & -38:08:22.98 & $Gaia$ DR2 \\ % \text{(deg:min:sec)}
        $\mu_{\text{R.A.}}$ & R.A. proper motion (mas y$^{-1}$) & -21.126 $\pm$ 0.046 & $Gaia$ DR2 \\
        $\mu_{\text{DEC}}$ & DEC proper motion (mas y$^{-1}$) & -0.988 $\pm$ 0.037 & $Gaia$ DR2 \\
        $\varpi$ & Parallax (mas) & 1.516 $\pm$ 0.035 & $Gaia$ DR2 \\
        \\
        Photometric Properties & & & \\
        V & (mag) & 12.704 $\pm$ 0.021 & APASS \\
        B & (mag) & 13.419 $\pm$ 0.010 & APASS \\
        J & (mag) & 11.481 $\pm$ 0.020 & 2MASS \\
        H & (mag) & 11.244 $\pm$ 0.030 & 2MASS \\
        K & (mag) & 11.141 $\pm$ 0.020 & 2MASS \\
        SDSS g & (mag) & 13.006 $\pm$ 0.022 & SDSS \\
        SDSS i & (mag) & 12.363 $\pm$ 0.027 & SDSS \\
        SDSS r & (mag) & 12.514 $\pm$ 0.065 & SDSS \\
        W1 & (mag) & 11.095 $\pm$ 0.030 & WISE \\
        W2 & (mag) & 11.148 $\pm$ 0.030 & WISE \\
        W3 & (mag) & 11.125 $\pm$ 0.107 & WISE \\
        NUV & (mag) & 18.471 $\pm$ 0.066 & GALEX \\
        \hline
        \end{tabular}
    \centering
    \caption{NGTS-13 (2MASS J11445767-38082292 / TIC 454069765) astrometric and photometric properties. The astrometric and photometric properties as well as the spectroscopically derived atmospheric parameters in Table \ref{tab:spectra_param} are used as inputs into our EXOFASTv2 global fit including the SED fit (see section \ref{sec:glob}).}
    \label{tab:star_param}
\end{table*}

\subsection{Spectral Analysis}

\begin{figure}
  \centering
  \includegraphics[width=0.99\columnwidth]{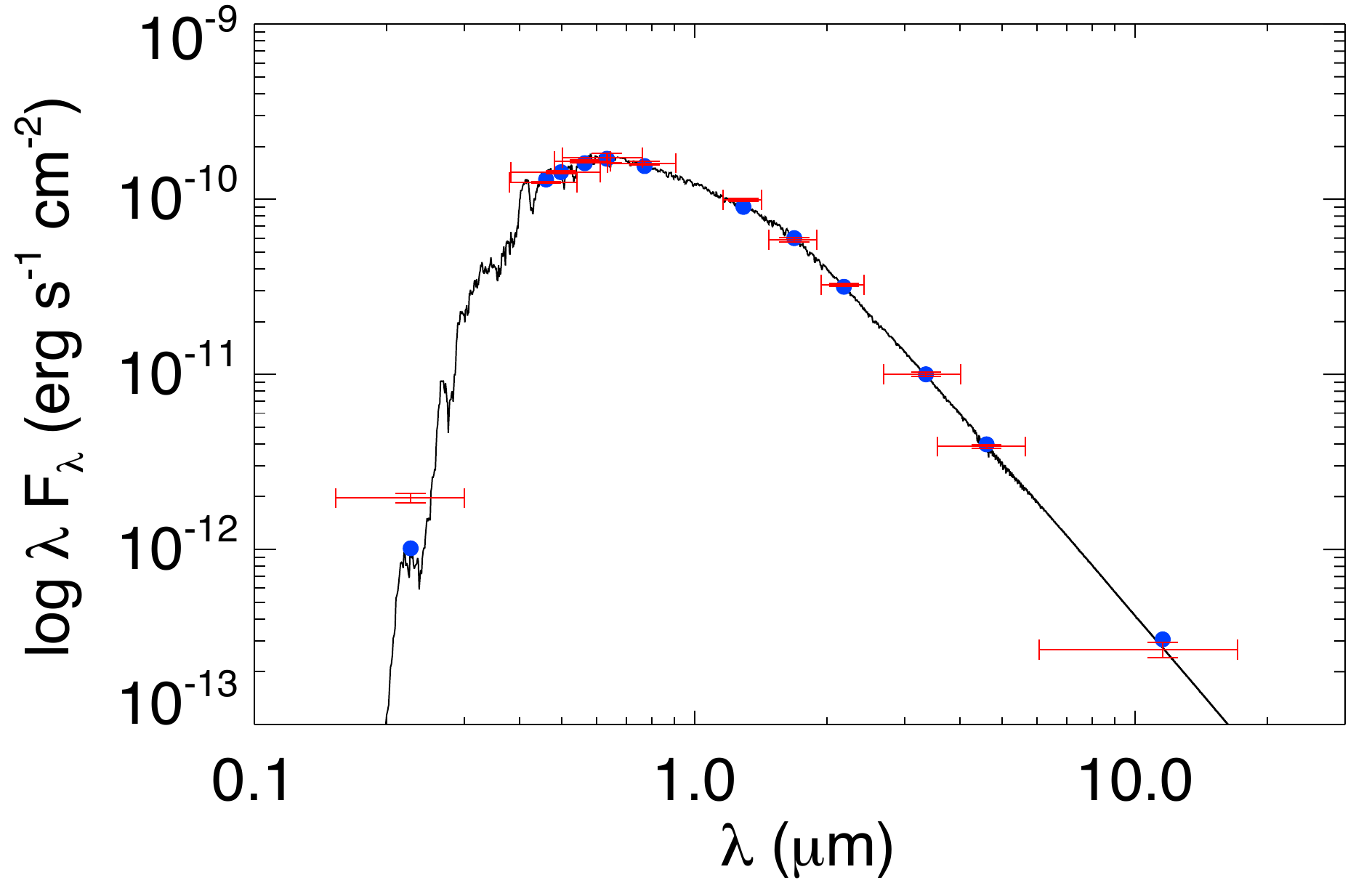}
  \caption{The Spectral Energy distribution model (black line) of NGTS-13 from the EXOFASTv2 fit with broad band averages (blue circles) and broad band measurements (red). The error bars in wavelength denote the bandwidth of the corresponding filter and the error bars in flux denote the measurement uncertainty.}
  \label{fig:sed}
\end{figure}

\label{sec:spectrased}
We used wavelet analysis to extract stellar atmospheric parameters from the co-added 13 CORALIE spectroscopic observations of NGTS-13 using the methodology set out in \citet{Gill2018,Gill2019}. With the wavelet analysis we find NGTS-13 to have surface rotational velocity \vsini = 6.2\,$\pm$\,1.2\,$\kms$. We also derived stellar atmospheric parameters using SpecMatch-Emp \citep{Yee2017}. SpecMatch-Emp uses a large library of stars with well-determined parameters to match the input spectra and derive spectral parameters. We use a spectral region that includes the Mg I b triplet (5100 - 5400 $\AA$) to match our spectra. SpecMatch-Emp uses $\chi^{2}$ minimisation and a weighted linear combination of the five best matching spectra in the SpecMatch-Emp library to determine $\teff$, $\logg$, and $\feh$. We combine the wavelet analysis results and SpecMatch-Emp results to create wide uncertainties that include the uncertainty range of both methods to use as priors for their final derivation as described below. We present these spectroscopically derived parameters from both methods and the adopted combined values in table \ref{tab:spectra_param}.

\subsection{Global Analysis}
\label{sec:glob}

We derive both planet and stellar parameters using EXOFASTv2 \citep{Eastman2013,Eastman2017,Eastman2019}, which can globally fit all photometry and RV data. A full description of EXOFASTv2 is given in \citet{Eastman2019}, which can fit any number of transits and RV sources while exploring the vast parameter space through a differential evolution Markov Chain coupled with a Metropolis-Hastings Monte Carlo sampler. A built-in Gelman-Rubin statistic \citep{GelmanRubin1992,Gelman2003,Ford2006} is used to check the convergence of the chains.

We use Modules for Experiments in Stellar Astrophysics (MESA) Isochrones \& Stellar Tracks \citep[MIST;][]{Choi2016,Dotter2016} isochrones and the spectral energy distribution (SED) within the EXOFASTv2 fit to determine host star parameters. For the SED fit within EXOFASTv2 \citep{StassunTorres2016} we use photometry from APASS DR9 BV \citep{Henden2016}, 2MASS JHK \citep{Skrutskie2006}, SDSS DR12 {\it gri} \citep{Alam2015}, ALL-WISE W1, W2 and W3 \citep{Wright2010}, and GALEX NUV \citep{Bianchi2011}, which are presented in Table \ref{tab:star_param}. The SED fit within EXOFASTv2 puts sytematic floors on the broadband photometry errors \citep{StassunTorres2016}. The EXOFASTv2 SED fit is presented in Figure \ref{fig:sed}. We use our spectroscopically derived $\teff$, $\logg$, and $\feh$ values as priors in the global model as well as a Gaussian prior for parallax from the value and uncertainty in $Gaia$ DR2. We put an upper limit on the extinction based on the maximum in line-of-sight value from the \citet{Schlafly2011} galaxy dust map. All other fitted and derived parameters from our EXOFASTv2 model have conservative physical boundaries that are detailed in Table 3 of \citet{Eastman2019}, which also gives a thorough explanation of each parameter.

Due to the long exposure time of the TESS full frame exposures we must account for smearing of the light curve. Therefore, within EXOFASTv2 we specify an exposure time of 30 minutes and average over 10 data points to integrate a model over the exposure time equivalent to a midpoint Riemann sum \citep{Eastman2019}. We assign the TESS FFI photometry to the built-in EXOFASTv2 "TESS" band and the NGTS photometry to the "R" band for the computation of the limb darkening coefficients. We found the EXOFASTv2 "R" band to have the most similar bandpass to NGTS in the EXOFASTv2 framework, which only allows filters defined in \citet{ClaretBloemen2011}. We tested the use of this filter by running a fit without TESS data and NGTS data assigned to the "TESS" filter; we obtained limb darkening parameters within uncertainties to those presented in Table \ref{tab:global_param}. We also ran a fit including TESS data assigned to the "TESS" filter and the NGTS data assigned to the "Kepler" filter; we again found limb darkening parameters within uncertainties to those presented. We account for dilution of the photometry lightcurves, which we describe in Section~\ref{sec:dil}. For our EXOFASVTv2 global model and Markov chain Monte Carlo (MCMC) fit we used 56 walkers (2 $\times$ n$_{\text{parameters}}$), or chains, and allowed the fit to run until convergence (132776 steps).

\begin{table*}
\centering
\begin{tabular}{cccTT}
\hline\hline
Parameter & Unit & \textbf{Value} & Value (High \msol) & Value (Low \msol) \\
 &  & \textbf{(Adopted Solution)} & (60\% Probability) & (40\% Probability) \\
\hline
\\
\textbf{Stellar Parameters} & & \\
$M_*$ & Mass (\msol)  & \boldmath{$1.30^{+0.11}_{-0.18}$} & $1.34^{+0.05}_{-0.05}$ & $1.16^{+0.05}_{-0.07}$ \\ 
$R_*$ & Radius (\rsol) & \boldmath{$1.788^{+0.057}_{-0.057}$} & $1.792^{+0.051}_{-0.049}$ & $1.787^{+0.052}_{-0.052}$ \\
$L_*$ & Luminosity (\lsol)  & \boldmath{$3.29^{+0.21}_{-0.21}$} & $3.34^{+0.19}_{-0.19}$ & $3.25^{+0.19}_{-0.20}$ \\
$\rho_*$ & Density (cgs)  & \boldmath{$0.31^{+0.04}_{-0.04}$} & $0.33^{+0.03}_{-0.03}$ & $0.28^{+0.03}_{-0.03}$ \\
$\logg$ & Surface gravity (cgs)  & \boldmath{$4.04^{+0.05}_{-0.05}$} & $4.06^{+0.02}_{-0.02}$ & $4.00^{+0.03}_{-0.03}$ \\
$\teff$ &Effective Temperature (K) & \boldmath{$5819^{+73}_{-73}$} & $5831^{+62}_{-72}$ & $5799^{+64}_{-73}$ \\
$\feh$ &Metallicity (dex) & \boldmath{$0.25^{+0.17}_{-0.17}$} & $0.28^{+0.13}_{-0.14}$ & $0.12^{+0.16}_{-0.18}$ \\
Age & Age (Gyr)  & \boldmath{$4.23^{+2.65}_{-1.59}$} & $3.68^{+0.58}_{-0.62}$ & $6.55^{+1.30}_{-0.89}$ \\
EEP & Equal Evolutionary Point & \boldmath{$413^{+47}_{-18}$} & $404^{+9}_{-14}$ & $453^{+4}_{-5}$ \\
A$_{V}$ & V-band extinction (mag) & \boldmath{$0.164^{+0.044}_{-0.044}$} & $0.159^{+0.036}_{-0.064}$ & $0.149^{+0.043}_{-0.071}$ \\
$d$ & Distance (pc) & \boldmath{$657^{+15}_{-15}$} & $659^{+15}_{-14}$ & $655^{+15}_{-14}$ \\
\\
\textbf{Planet Parameters} & & \\
$M_P$ & Mass (\mj)  & \boldmath{$4.84^{+0.44}_{-0.44}$} & $5.05^{+0.26}_{-0.26}$ & $4.54^{+0.26}_{-0.29}$ \\
$R_P$ & Radius (\rj) & \boldmath{$1.142^{+0.046}_{-0.046}$} & $1.142^{+0.036}_{-0.034}$ & $1.143^{+0.040}_{-0.038}$ \\
$P$ & Period (days) & \boldmath{$4.119027^{+0.000023}_{-0.000023}$} & $4.119027^{+0.000020}_{-0.000020}$ & $4.119027^{+0.000020}_{-0.000020}$ \\
$T_C$ & Time of conjunction (\bjdtdb) & \boldmath{$2458572.8247^{+0.0018}_{-0.0018}$} & $2458572.8247^{+0.0015}_{-0.0015}$ & $2458572.8246^{+0.0015}_{-0.0015}$ \\
$a$ & Semi-major axis (AU) & \boldmath{$0.0549^{+0.0015}_{-0.0025}$} & $0.0556^{+0.0007}_{-0.0007}$ & $0.0529^{+0.0007}_{-0.0011}$ \\
$i$ & Inclination (Degrees) & \boldmath{$88.7^{+1.2}_{-1.2}$} & $88.8^{+0.8}_{-1.1}$ & $87.9^{+1.4}_{-1.3}$ \\
$e$ & Eccentricity & \boldmath{$0.086^{+0.034}_{-0.034}$} & $0.091^{+0.023}_{-0.022}$ & $0.080^{+0.023}_{-0.022}$ \\
$\omega_*$ & Argument of Periastron (Degrees) & \boldmath{$-146.4^{+24.2}_{-24.2}$} & $-143.0^{+13.0}_{-18.0}$ & $-161.0^{+24.0}_{-29.0}$ \\
$K$ & RV semi-amplitude (m/s) & \boldmath{$526.1^{+29.5}_{-29.5}$} & $526.0^{+24.0}_{-24.0}$ & $525.0^{+25.0}_{-27.0}$ \\
$\delta$ & Transit depth (fraction) & \boldmath{$0.0043^{+0.0004}_{-0.0004}$} & $0.0043^{+0.0001}_{-0.0001}$ & $0.0043^{+0.0001}_{-0.0001}$ \\
$\tau$ & Ingress/egress transit duration (days) & \boldmath{$0.0150^{+0.0022}_{-0.0022}$} & $0.0137^{+0.0009}_{-0.0003}$ & $0.0143^{+0.0018}_{-0.0008}$ \\
$T_{14}$ & Total transit duration (days) & \boldmath{$0.2176^{+0.0030}_{-0.0030}$} & $0.2173^{+0.0021}_{-0.0020}$ & $0.2181^{+0.0024}_{-0.0022}$ \\
$b$ & Transit Impact parameter & \boldmath{$0.156^{+0.145}_{-0.145}$} & $0.150^{+0.130}_{-0.100}$ & $0.240^{+0.150}_{-0.160}$ \\
$\rho_P$ & Density (cgs) & \boldmath{$4.02^{+0.55}_{-0.55}$} & $4.20^{+0.43}_{-0.41}$ & $3.77^{+0.46}_{-0.43}$ \\
log g$_P$ & Surface gravity (cgs) & \boldmath{$3.96^{+0.06}_{-0.06}$} & $3.98^{+0.03}_{-0.03}$ & $3.94^{+0.04}_{-0.04}$ \\
$T_{eq}$ & Equilibrium temperature (K) & \boldmath{$1605^{+30}_{-30}$} & $1595^{+21}_{-22}$ & $1627^{+24}_{-26}$ \\
$S$ & Insolation Flux ($S_{\oplus}$) & \boldmath{$1096^{+90}_{-90}$} & $1064^{+58}_{-59}$ & $1153^{+70}_{-70}$ \\
\hline
\end{tabular}
\\
\begin{tabular}{llcc}
\\
Wavelength Parameters & & R (NGTS) & TESS\\ 
$u_{1}$ & linear limb-darkening coeff & $0.314^{+0.047}_{-0.047}$ & $0.331^{+0.049}_{-0.050}$ \\ 
$u_{2}$ & quadratic limb-darkening coeff & $0.248^{+0.049}_{-0.049}$ & $0.307^{+0.049}_{-0.049}$ \\ 
\end{tabular}
\\
\begin{tabular}{llcc}
\\ 
Transit Parameters & & NGTS & TESS \\ 
$\sigma^{2}$ & Added Variance & $0.00000501^{+0.00000021}_{-0.00000020}$ & $0.00000034^{+0.00000003}_{-0.00000003}$  \\ 
$F_0$ & Baseline flux & $1.000270^{+0.000047}_{-0.000047}$ & $0.999972^{+0.000026}_{-0.000026}$  \\ 
\end{tabular}
\\
\begin{tabular}{llc}
 \\ 
RV Parameters &  & CORALIE \\ 
$\gamma_{\rm rel}$ & Relative RV Offset (m/s) & $27029^{+17}_{-17}$  \\ 
$\sigma_J$ & RV Jitter (m/s) & $0^{+40}_{-0}$ \\ 
$\sigma_J^2$ & RV Jitter Variance & $-800^{+2400}_{-1100}$ \\ 
o$-$c rms & Root Mean Square of residuals (\ms) & 40.6 \\ 
\\
\hline
\end{tabular}
\caption{NGTS-13 stellar and companion parameters derived from our EXOFASTv2 modeling and analysis described in Section \ref{sec:analysis}. We present two solution split between the peaks of our bimodal stellar mass posterior distribution, as described in Section \ref{sec:bimodal}. The final adopted solution is in bold and uses the median value and upper and lower boundaries of the 68\% confidence interval we computed, as described in Section \ref{sec:bimodal}. For the wavelength, transit, and RV parameters listed in the lower part of the table we found the two solutions to be very similar and well below uncertainties and only present the median values of all the MCMC steps. The boundaries (besides the specific priors mentioned in Section \ref{sec:analysis}) and thorough explanations of all fitted and derived parameters from EXOFASTv2 are detailed in Table 3 of \citet{Eastman2019}.}
\label{tab:global_param}
\end{table*}

\subsubsection{Bimodal Stellar Mass and Age}
\label{sec:bimodal}

\begin{figure}
  \centering
   \includegraphics[width=0.96\columnwidth]{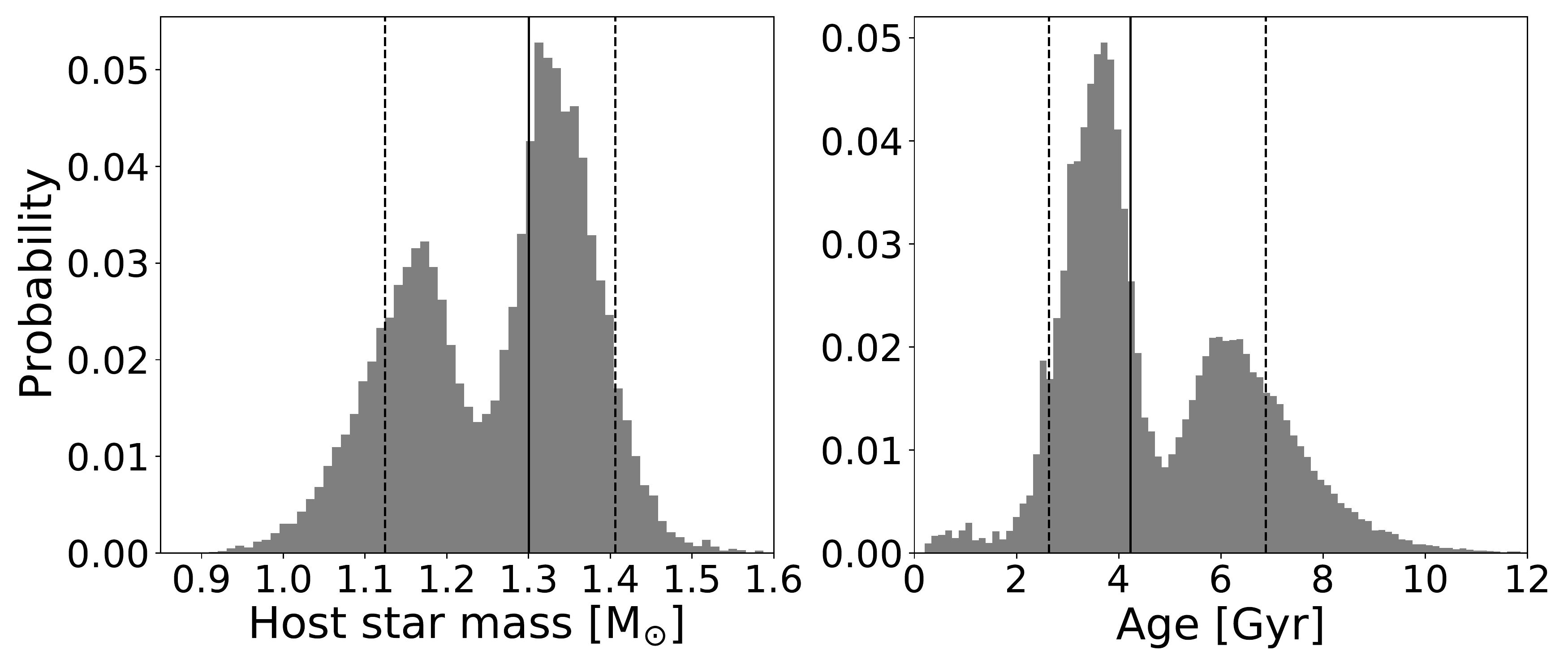}
   \caption{Stellar mass and age posterior distributions from EXOFASTv2. These distributions show the relative probability between the two peaks of the bimodal distributions, which is approximately 60\% to 40\% in favor of a more massive, younger system. We split the stellar mass distribution in the valley at 1.25 M$_{\odot}$ to create two solutions presented in table \ref{tab:global_param}. 
   For our final solution we adopt the median value (vertical black lines) and upper and lower boundaries (dashed vertical lines) of the 68\% confidence intervals we computed, as described in Section \ref{sec:bimodal}.}
  \label{fig:agemasspost}
\end{figure}

We inspected the posterior distributions of each fitted and derived parameter and found bimodal distributions in both the stellar mass and age. This type of bimodality between two different stellar ages and masses occurs due to a degeneracy between two MIST isochrone tracks that EXOFASTv2 interpolates between and has been seen in several recent studies \citep[e.g.,][]{Ikwut-Ukwa2020,Pepper2020,Carmichael2020a,Carmichael2020b}. Similar to those studies we split the host star mass posterior distribution at the valley between the two peaks, or 1.25 M$_{\odot}$, and extract two separate solutions, both of which are presented in Table \ref{tab:star_param}. We show the stellar mass and age posterior distributions in Figure \ref{fig:agemasspost}. The higher mass solution is moderately more probable with a probability of 60\% compared to 40\% for the lower mass solution.

However, for our final adopted solution instead of picking one peak we define wider uncertainties by computing confidence intervals for each posterior distribution based on the smallest area in the posterior distribution function that contains 68\% of the MCMC steps. We compute these confidence intervals using a subroutine within the open source python package MC3\footnote{https://mc3.readthedocs.io/en/latest/index.html} \citep{Cubillos2017}. In summary the code orders the posterior distribution by value and creates a smoothed posterior density distribution (PDF), computes a cumulative distribution function (CDF) of the ordered PDF, and computes where the CDF reaches 68\% of its maximum value. The minimum and maximum parameter values inside this interval are the boundaries of the 1$\sigma$ confidence interval. We use the median value within the confidence interval as our final value for each parameter and the minimum and maximum values within the confidence interval as their uncertainties, which are in bold in Table \ref{tab:global_param}. Figure \ref{fig:agemasspost} displays these values for the host star mass and age.

\subsubsection{Dilution} \label{sec:dil}

NGTS-13 has several close companions including two stars only slightly dimmer: TIC 454069767 (TESS mag = 14.1, 22.5'' west) and TIC 454069772 (TESS mag = 13.4, 57'' south-east), see Figure~\ref{fig:tesscut}. We expect the TESS photometry of NGTS-13 to be completely diluted by the companion 22.5'' away and partially diluted by the companion 57.3'' away. One other close companion has a TESS mag $<$ 18, while all other companions are dimmer and not expected to dilute at all. 

To account for dilution we calculate a dilution factor, given as a contrast ratio $C$, between the combined flux of all contaminating sources and the flux of the target star:
\begin{equation}
    C = F_{cont}/F_{target}
\end{equation}
where $F_{target}$ and $F_{cont}$ are the flux contributions within the photometric aperture from the target and contaminating stars respectively. We follow the method detailed in \citet{Bryant2020} to estimate dilution. 

We determine the point-spread function (PSF) in the region of the camera surrounding NGTS-13 and extract 2D Gaussian shape parameters \citep{Pal2009} using the Cluster Difference Imaging Photometric Survey \citep[CDIPS;][]{Bouma2019}. For NGTS-13 we find shape parameters of $S$ = 1.88\,$\pm$\,0.047 (FHWM = 1.71 pixels), $D$ = 0.033\,$\pm$\,0.011, and $K$ = 0.169\,$\pm$\,0.007. We consider all the stars in TICv8 \citep{Stassun2019} within 1' of NGTS-13 and find a dilution factor of 19.10$^{+0.05}_{-0.04}$\% in the TESS photometry. We apply \citet{Bryant2020}'s method to our NGTS data assuming a circular Gaussian PSF ($D$ = $K$ = 0) and a FWHM = 2\,$\pm$\,0.25 pixels, so $S$ = 1.38\,$\pm$\,0.3. We determine a dilution factor of 0.64$^{+0.33}_{-0.21}$\% for our NGTS photometry. Both of these dilution factors are used as inputs into the EXOFASTv2 global model. We note that EXOFASTv2 defines their dilution factor as the fractional contribution from neighboring stars, or $C$/(1+$C$).

%--------------------------------------------------------------------
\section{Discussion}\label{sec:discussion}

Our analysis finds NGTS-13b to have a mass of  4.84\,$\pm$\,0.44\,\mjup and radius of  1.142\,$\pm$\,0.046\,\rjup at a short 4.12 day orbital period yielding an equilibrium temperature of 1605\,$\pm$\,30\,K, making NGTS-13b a massive hot Jupiter. Although close-in very massive planets are the easiest to detect, they are far from the most common. As of 16 September 2020 the NASA Exoplanet Archive\footnote{\label{refnote1}\url{http://exoplanetarchive.ipac.caltech.edu}} lists 4277 confirmed planets; however, only 1764 have measured masses and only 1691 of these are below 13\,\mjup (the approximate border between planets and brown dwarfs). Of planets with measured masses 971 are giant ($M_{P}$ = 0.5 - 13\,\mjup) with 223 of them having masses greater than 4\,\mjup. Here we only focus on planets that have well determined masses and radii (percentage errors < 10\%), which consists of 225 planets, of which 155 are giants including only 17 massive giants ($M_{P}$ > 4\,\mjup). These very massive planets have a similarly low occurrence rate to the even more massive brown dwarfs, which are indeed known to be rare \citep[the brown dwarf desert, e.g.,][]{MarcyButler2000,Grether2006}. 

\begin{figure}
  \centering
   \includegraphics[width=0.99\columnwidth]{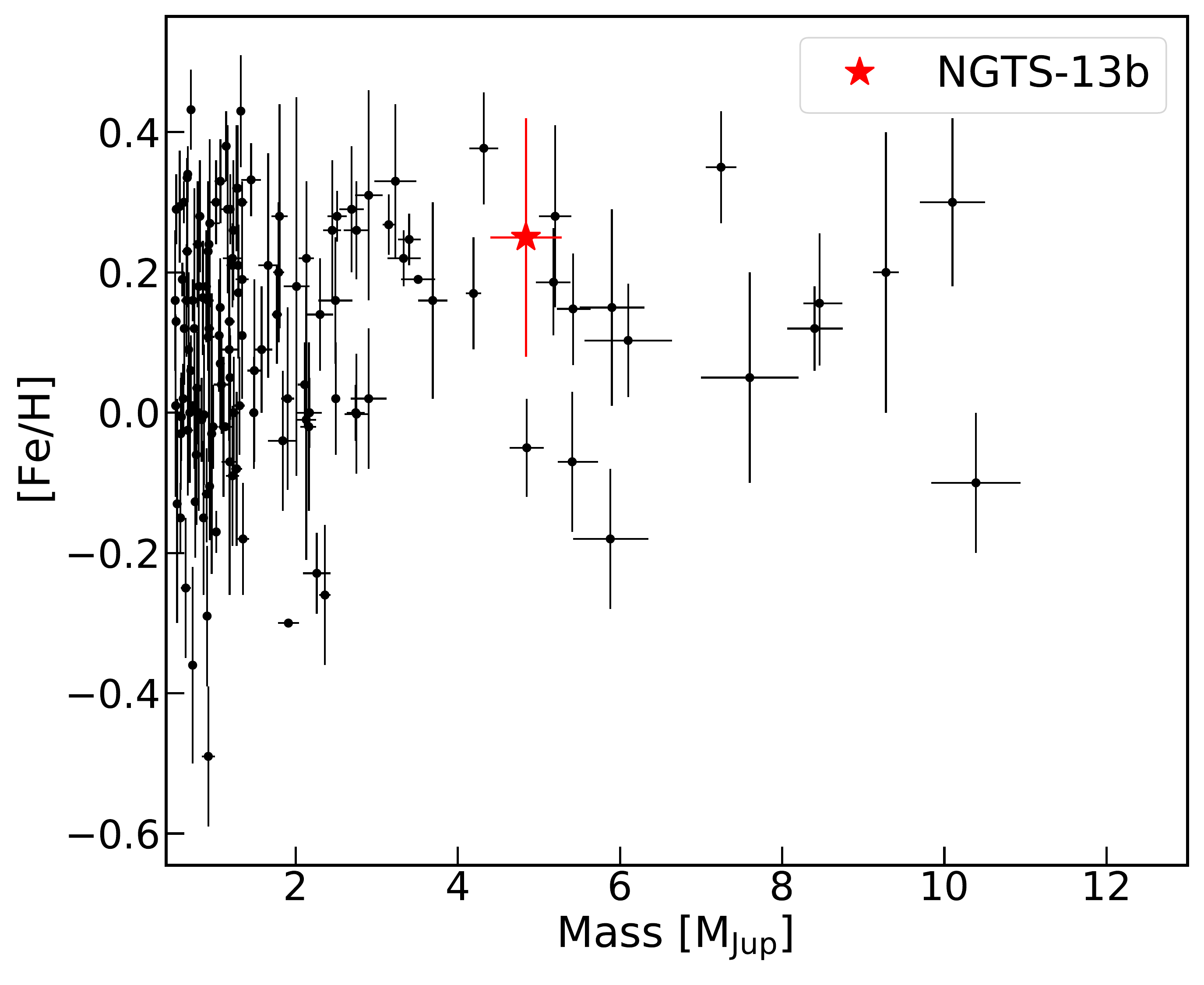}
   \caption{[Fe/H] and mass relationship for giant planets (0.5-13\,\mjup) with percentage errors $<$10\% for both mass and radius and a defined [Fe/H] in the NASA Exoplanet Archive\footref{refnote1}, which consists of 137 previously known planets including 17 with masses greater than 4\,$\mjup$.}
  \label{fig:fehmass}
\end{figure}

\begin{figure*}
  \centering
    \includegraphics[width=1.21\columnwidth]{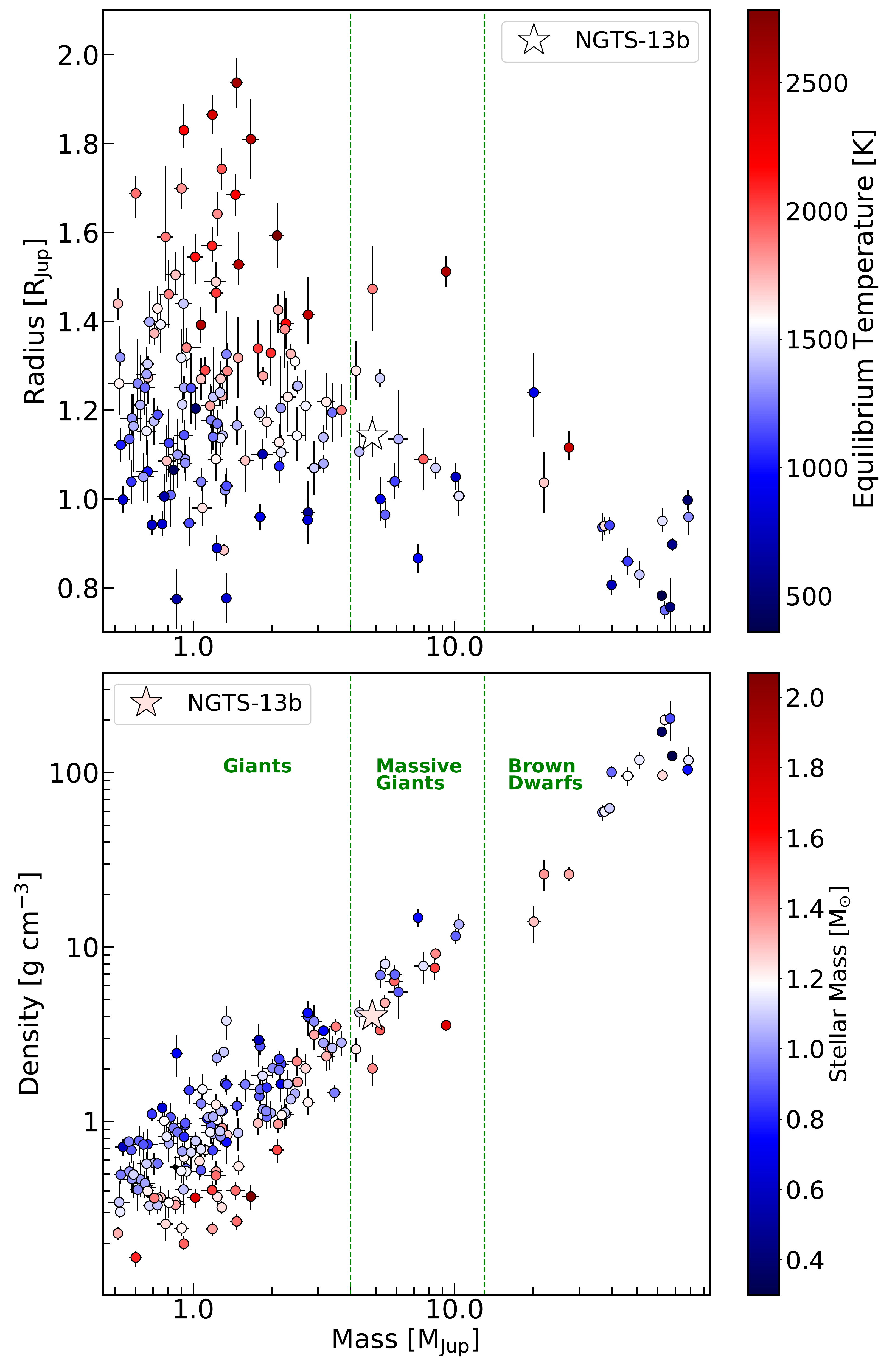}
   \caption[]{Radius (\textit{top}) and density (\textit{bottom}) as a function of mass for transiting brown dwarfs and giant planets (0.5-13\,\mjup) with percentage errors $<$10\% for both mass and radius in the NASA Exoplanet Archive\footref{refnote1}. Transiting brown dwarfs are from the literature summary found in \citep{Mireles2020} for a total of 16 with well-defined masses and radii. The stars display NGTS-13b and the vertical dashed lines highlight 4\,$\mjup$ as well as 13\,$\mjup$ (the approximate starting mass of brown dwarfs). The colors of the symbols correspond to equilibrium temperature in the top plot and host star mass in the bottom plot. The downward trend in maximum radii along increasing mass exhibits the effect of stronger surface gravity in more massive objects, which is clearly displayed by the linear trend in density. In the top plot we only display planets that also have an equilibrium temperature in the Archive, which consists of 133 previously known planets including 14 with masses greater than 4\,$\mjup$.}
  \label{fig:mrp}
\end{figure*}

\subsection{Massive substellar companion populations}

Previous studies have suggested that 4\,$\mjup$ could be an approximate border for a distinct massive giant planet population compared to lower-mass giant planets. \citet{Santos2017} find that massive giant planets tend to orbit stars with $\feh$ distributions statistically similar to field stars, unlike lower-mass giant planets which occur more frequently around metal-rich stars. This could be evidence for two different formation mechanisms where lower-mass giants may form via core accretion \citep{Pollack1996} and higher mass giants may form via disk instability \citep{Boss1997}, as formation via disk instability is not as metallicity dependent as core-accretion
\citep{Boss2002,Cai2006}.

Additionally, \citet{Schlaufman2018} find that substellar companions with masses $\lesssim$\,4\,\mjup preferentially orbit metal rich stars (a property associated with core accretion), while companions with masses $\gtrsim$\,10\,\mjup do no share this property. \citet{Maldonado2019} also find that massive giant planets tend to have host stars with lower metallicities than lower-mass giant planets and conclude that the core-accretion planet formation mechanism achieves its maximum efficiency for planets with masses in the range 0.2 – 2\,\mjup. \citet{GodaMatsuo2019} conclude that the mean metallicity of stars hosting companions between 4 and 25\,\mjup is lower than that of the sample of companions with masses 0.3 to 4\,\mjup, and the mean metallicity of stars with companions more massive than 25\,$\mjup$ is much lower than those of the other two sub-samples. \citet{GodaMatsuo2019} use 25\,\mjup as the lower mass limit of brown dwarfs (instead of the deuterium burning limit) because they found the mean metallicity of G-type host stars with objects lighter than 25\,\mjup to be much higher than those with objects more massive than 25\,\mjup; they suggest 25\,\mjup is the upper mass limit for core-accreted planets. However, other previous studies suggest lower mass brown dwarfs (13 - 42.5\,\mjup) form via disk instability, e.g., \citet{MaGe2014}.

\citet{Adibekyan2019} analyzes the known giant planet population and his results do not support that 4\,\mjup is a transition point between two separate formation channels, but his results do suggest that high mass planets can form through different mechanisms depending on their initial environment. With a metallicity of $\feh$ = 0.25\,$\pm$\,0.17 NGTS-13b's host star is metal-rich and does not provide additional evidence that more massive-giant planets prefer a different formation mechanism than core accretion. We put NGTS-13b in this context in Figure~\ref{fig:fehmass} which shows the metallicity of giant planets from 0.5 - 13\,$\mjup$. 

\subsection{NGTS-13b in context of the mass-radius relationship}

We put NGTS-13b in terms of its density, $\rho_P$ = 4.02\,$\pm$\,0.55\,g\,cm$^{-3}$, in Figure~\ref{fig:mrp} by displaying it in the mass-radius diagram with other giant planets and brown dwarfs as well as density as a function of mass. We color symbols in the mass-radius diagram based on equilibrium temperature to show the effect high equilibrium temperatures can have on radii. We can see in Figure~\ref{fig:mrp} the larger maximum radii for planets with masses lower than $\sim$4\,$\mjup$, but this is explained by the redder coloring of these symbols indicating higher equilibrium temperatures (resulting in more inflation). NGTS-13b has a typical radius compared to planets with similar masses suggesting that even with its high equilibrium temperature it is not significantly affected by bloating (inflated radius due to a planet's atmosphere being heated from stellar irradiation and expanding), which is expected for massive planets \citep{Sestovic2018}. The downward trend in radii (and increasing trend in density) with larger mass in Figure~\ref{fig:mrp} highlights the importance of the larger surface gravity in more massive objects.

\subsection{A slightly evolved host star}

NGTS-13 has an effective temperature of $T_{\rm eff}$ = 5819\,$\pm$\,73\,K, suggesting it is a G2 type star \citep{PecautMamajek2013}. However, the rather large stellar radius of $R_*$ = 1.79\,$\pm$\,0.06\,$\rsol$ indicates NGTS-13 has evolved slightly as this radius is larger than expected for a main sequence star of this temperature, and its surface gravity \logg = 4.04\,$\pm$\,0.05 suggests the star is a subgiant (\logg $<$ 4.1). Our MIST isochrone analysis (see Section~\ref{sec:glob}) finds an equivalent evolutionary point (EEP) value of 413$^{+47}_{-18}$, which still puts NGTS-13 in the main sequence evolutionary phase, as the terminal age main sequence EEP does not start until EEP = 454, see \citet{Choi2016}, \citet{Dotter2016}, and the MIST documentation\footnote{http://waps.cfa.harvard.edu/MIST/README\_tables.pdf}. Additionally, it is less likely in terms of evolutionary timescales to observe a star near or past the turnoff point than to observe it in the middle of the main sequence, therefore NGTS-13 is likely still fusing Hydrogen in its core but has likely begun the transition from the main sequence to the red giant branch placing it in the subgiant branch.

\citet{Lillo-Box2016} found a relative lack of hot Jupiters around giant and subgiant stars, suggesting that the close-in massive planets around main-sequence stars are engulfed by the star as it evolves. However, other studies do not find a lack of hot Jupiters around evolved hosts; additionally NGTS-13 is likely still burning hydrogen in its core and has not evolved long enough to engulf planets near the orbital distance of NGTS-13b. \citet{Grunblatt2019} found that low-luminosity red giant branch evolved stars have a tentatively higher population of close-in giant ($R_{P}$ $>$\,1\,\rjup) planets than main-sequence stars suggesting that stellar evolution does not affect close-in giant planet occurrence significantly until stars are substantially in the red giant branch with radii of 5-6\,\rsol. \citet{Zhou2019} studied the occurrence rate of hot Jupiters as a function of stellar mass and did not find any statistically consistent trends with stellar mass and that hot Jupiters are just as abundant around main-sequence A stars as they are around F and G stars.

The \mstar = 1.30$^{+0.11}_{-0.18}$\,\msol stellar mass of NGTS-13 is larger than a typical G-type main-sequence star, which suggests it evolved to a cooler temperature and is a "retired" F star. A relatively larger host star mass is consistent with previous studies that found massive planets to be more common around massive stars \citep{Santos2017,Maldonado2019}. \citet{Santos2017} find that more massive giant planets are more common around more massive hosts, which are often more evolved; they suggest the two distinct populations of giant planets could be related to evolved stars not showing a clear metallicity-giant planet correlation. 

We display host star mass by color in Figure~\ref{fig:mrp}. Only one host star has a mass below 0.9\,$\msol$ in our sample of 17 well-defined massive giant planets. \citet{Grunblatt2018} found that close-in giant planets around evolved stars tend to have more eccentric orbits than those around main sequence stars, but they focus on giant stars and orbital periods longer than 4.5 days. We explore NGTS-13b's eccentricity further in Section~\ref{sec:ecc}.

\subsection{A slightly eccentric orbit} \label{sec:ecc}

\begin{figure}
  \centering
   \includegraphics[width=0.9\columnwidth]{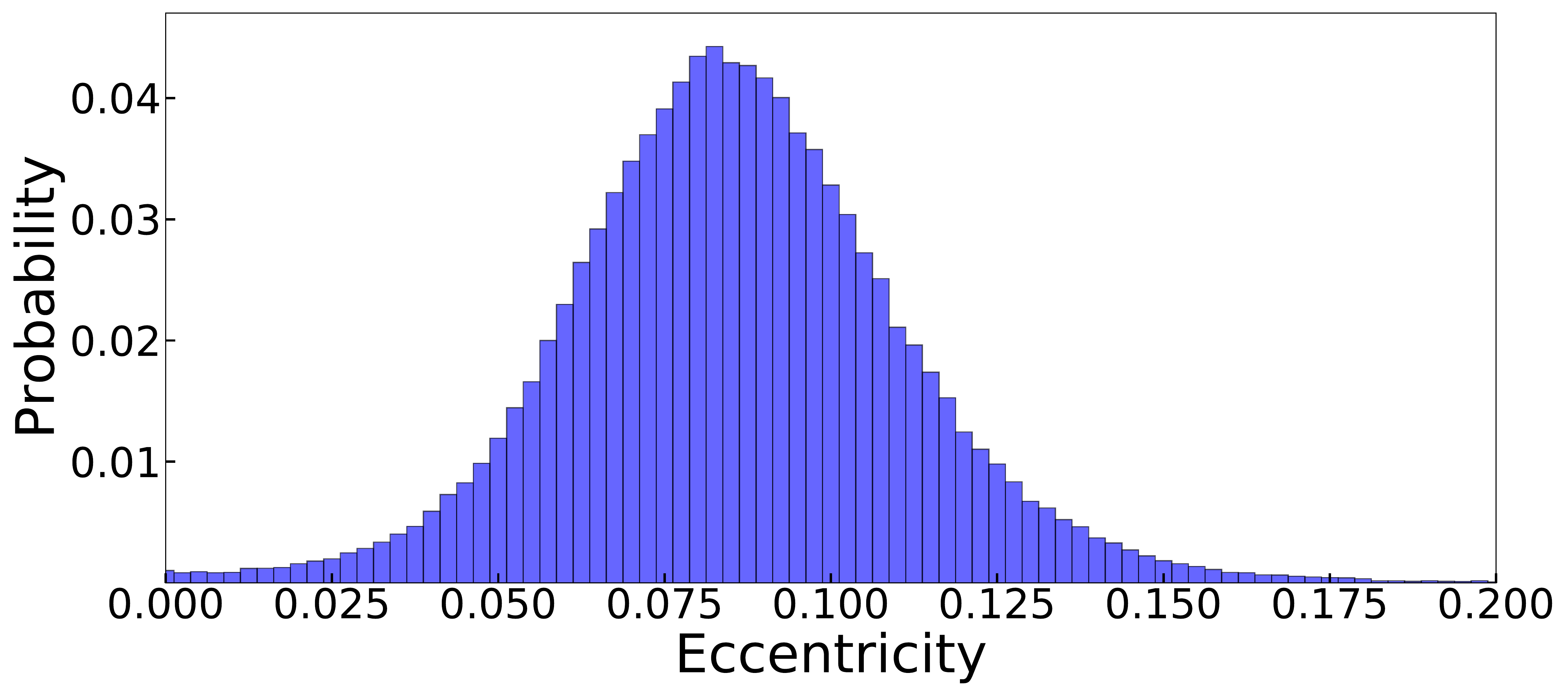}
   \caption{Posterior distribution of the eccentricity from our EXOFASTv2 Markov Chain Monte Carlo (MCMC) analysis including all chains and steps of the distribution.}
  \label{fig:ecc}
\end{figure} 

\begin{figure}
  \centering
   \includegraphics[width=0.99\columnwidth]{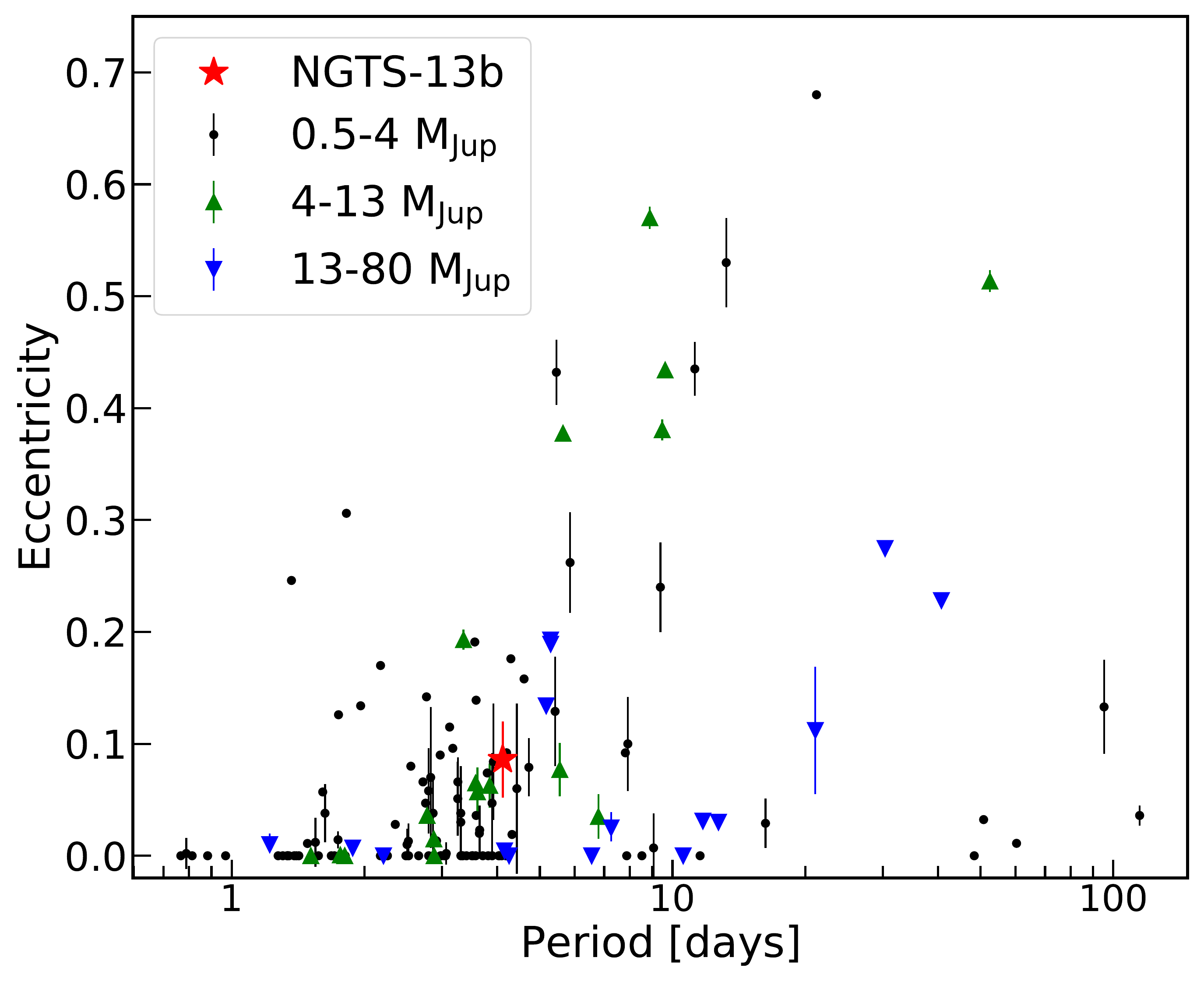}
   \caption{
   Eccentricity-period relationship for transiting brown dwarfs and giant planets (0.5-13\,$\mjup$) with percentage errors $<$10\% for both mass and radius and a defined eccentricity in the NASA Exoplanet Archive\footref{refnote1}, which consists of 129 previously known planets including 17 with masses greater than 4\,$\mjup$. Transiting brown dwarfs are from the literature summary found in \citep{Mireles2020} for a total of 16 with well-defined masses and radii.}
  \label{fig:massecc}
\end{figure}

NGTS-13b has a slightly eccentric, $e$ = 0.086\,$\pm$\,0.034, orbit with a short period of 4.12 days. Analyzing the posterior distribution of the eccentricity displayed in Figure~\ref{fig:ecc} we see that it is not centered around 0. We performed the test of \citet{Lucy1971}, and find the statistical significance of the eccentric fit to be P$(e > 0)$ = 0.9591, which just passes the 5\% significance level suggested by \citet{Lucy1971}. 

To further test the eccentric model we fit a Keplerian to the 13 RVs with priors on the time of conjunction and period and compare fits with and without forcing $\sqrt{e} \cos \omega_{*}$ and $\sqrt{e} \sin \omega_{*}$ to 0. We also allow an RV jitter term to vary, which corresponds to $k$ = 7 free parameters for the eccentric fit and $k$ = 5 for the circular fit. We determine the Bayesian Information Criterion (BIC), where given the free parameters $k$, number of measurements $n$, and maximized value of the likelihood function of the model $\mathscr{L}$:
\begin{equation}
    \text{BIC} = k \ln n - 2 \ln \mathscr{L}.
\end{equation}
When picking from several models the one with the lowest BIC is preferred. We find a $\Delta$BIC = 7.17 showing the eccentricity model is moderately favored over the circularized model. For this test we found an RV jitter of 31 \ms for the circular fit but 0 \ms for the eccentric fit as no additional RV uncertainty is needed for the eccentric model.

We note that for our final EXOFASTv2 analysis presented the model has a RV jitter variance $\sigma_J^2$ =  $-800^{+2400}_{-1100}$ which essentially reduces the error size of the RVs when evaluating the model. As noted in Equation 13 of \citet{Hara2019}, the uncertainty in the eccentricity is approximately proportional to the error of the RV measurement; therefore, this added RV jitter variance may cause an underestimation of our eccentricity uncertainties. \citet{Hara2019} show that the eccentricity estimate can be affected by an undetected correlated signal. Given the amplitude of the RV signal of NGTS-13b, we study the possibility that an undetected outer planetary companion would cause a nonzero eccentricity. We test this by adding a linear drift to our global EXOFASTv2 model. When allowing for an unconstrained linear RV drift we find a slope of only 1.1$^{+1.2}_{-1.4}$ {\mbox{m\,s$^{-1}$\,day$^{-1}$}\xspace} and an eccentricity $e$ = 0.084$^{+0.023}_{-0.021}$, similar to our presented model with no RV drift. The similar eccentricity suggests that NGTS-13b's eccentricity measurement is not caused by the model attempting to account for an undetected planet.

The slight eccentricity of NGTS-13b is not uncommon for massive planets (see Figure~\ref{fig:massecc}). \citet{Hansen2010} found that hot giant planets more massive than 3\,\mjup have the upper envelope of their eccentricity distribution shifted to lower periods compared to less massive hot Jupiters. Also \citet{DawsonJohnson2018} found that most hot Jupiters with periods $<$\,3 days have circular orbits, but some hot-Jupiters in the 3-10 day orbital period range occupy moderately eccentric orbits. Assuming a general tidal circularization effect for close-in planets from equation 3 of \citet{Adams2006}, NGTS-13b has a circularization timescale of 0.5 Gyr for a tidal quality factor $Q_{p}$ = 10$^{5}$, 5.0 Gyr for $Q_{p}$ = 10$^{6}$, and 50.3 Gyr for $Q_{p}$ = 10$^{7}$. Given the 4.23$^{+2.65}_{-1.59}$ Gyr age of the system, the current eccentricity may be an indication that NGTS-13b underwent dynamical interactions with other components in the system during its migration history, e.g.,  high-eccentricity tidal migration (see Section~3.1 of \citet{DawsonJohnson2018}). We put NGTS-13b's eccentricity into context with other giant planets and transiting brown dwarfs in Figure~\ref{fig:massecc}.

%--------------------------------------------------------------------
\section{Conclusions}\label{sec:conc}

We report the discovery of NGTS-13b, a massive hot Jupiter orbiting a subgiant star at a 4.12 day period. NGTS-13b has a mass of 4.84\,$\pm$\,0.44\,\mjup, radius of 1.142\,$\pm$\,0.046\,\rjup and eccentricity $e$ = 0.086\,$\pm$\,0.034 with an estimated equilibrium temperature of 1605\,$\pm$\,30\,K. Even with this high equilibrium temperature NGTS-13b is a dense object with $\rho_P$ = 4.02\,$\pm$\,0.55\,g\,cm$^{-3}$ as is expected for massive planets. The effective temperature, $T_{\rm eff}$ = 5819\,$\pm$\,73\,K, of NGTS-13 suggests it is a G-type star; however, its mass of \mstar = 1.30$^{+0.11}_{-0.18}$\,\msol and radius \rstar = 1.788\,$\pm$\,0.057\,\rsol suggest it is a "retired" F star in the subgiant branch.

NGTS-13b is a valuable addition to the relatively rare group ($<$20 currently known) of well-characterized (mass and radius percent errors $<$ 10\%) massive ($M_P$ $>$ 4\,$\mjup$) giant planets that are needed to understand the differences between giants, massive giants, and brown dwarfs. Previous studies have considered 4\,$\mjup$ as a boundary between core accretion planet formation and other formation mechanisms such as disk instability, based on metallicity distributions. However, NGTS-13 has a metallicity $\feh$ = 0.25\,$\pm$\,0.17 and does not give further evidence that these massive giants form through other mechanisms than core-accretion. 

%--------------------------------------------------------------------
\begin{acknowledgements}
      We thank the Swiss National Science Foundation (SNSF) and the Geneva University for their continuous support to our planet search programs. This work was carried out in the frame of the National Centre for Competence in Research $PlanetS$ supported by the Swiss National Science Foundation (SNSF). This work uses data collected under the NGTS project at the ESO La Silla Paranal Observatory. The NGTS facility is operated by the consortium institutes with support from the UK Science and Technology Facilities Council (STFC) under projects ST/M001962/1 and ST/S002642/1. This publication makes use of The Data \& Analysis Center for Exoplanets (DACE), which is a facility based at the University of Geneva (CH) dedicated to extrasolar planets data visualisation, exchange and analysis. DACE is a platform of the Swiss National Centre of Competence in Research (NCCR) PlanetS, federating the Swiss expertise in Exoplanet research. The DACE platform is available at https://dace.unige.ch. This paper includes data collected by the TESS mission. Funding for the TESS mission is provided by the NASA Explorer Program. This work has made use of data from the European Space Agency (ESA) mission Gaia (https://www.cosmos.esa.int/gaia), processed by the Gaia Data Processing and Analysis Consortium (DPAC, https://www.cosmos. esa.int/web/gaia/dpac/consortium). Funding for the DPAC has been provided by national institutions, in particular the institutions participating in the Gaia Multilateral Agreement. 
      JSJ acknowledges support by FONDECYT grant 1201371 and partial support from CONICYT project Basal AFB-170002.
      JVS acknowledges funding from the European Research Council (ERC) under the European Union’s Horizon 2020 research and innovation programme (project Four Aces; grant agreement No. 724427). MNG acknowledges support from MIT's Kavli Institute as a Juan Carlos Torres Fellow. Contributions by authors from the University of Warwick were supported by STFC consolidated grants ST/P000495/1 and ST/T000406/1.
      EG gratefully acknowledges support from the David and Claudia Harding Foundation in the form of a Winton Exoplanet Fellowship. The authors thank Aaron Dotter for his helpful advice.
\end{acknowledgements}

\bibliographystyle{aa}
\bibliography{bib}

\begin{thebibliography}{74}
\expandafter\ifx\csname natexlab\endcsname\relax\def\natexlab#1{#1}\fi

\bibitem[{{Adams} \& {Laughlin}(2006)}]{Adams2006}
{Adams}, F.~C. \& {Laughlin}, G. 2006, \apj, 649, 1004

\bibitem[{{Adibekyan}(2019)}]{Adibekyan2019}
{Adibekyan}, V. 2019, Geosciences, 9, 105

\bibitem[{{Alam} {et~al.}(2015){Alam}, {Albareti}, {Allende Prieto}, {Anders},
  {Anderson}, {Anderton}, {Andrews}, {Armengaud}, {Aubourg}, {Bailey}, {Basu},
  {Bautista}, {Beaton}, {Beers}, {Bender}, {Berlind}, {Beutler}, {Bhardwaj},
  {Bird}, {Bizyaev}, {Blake}, {Blanton}, {Blomqvist}, {Bochanski}, {Bolton},
  {Bovy}, {Shelden Bradley}, {Brandt}, {Brauer}, {Brinkmann}, {Brown},
  {Brownstein}, {Burden}, {Burtin}, {Busca}, {Cai}, {Capozzi}, {Carnero
  Rosell}, {Carr}, {Carrera}, {Chambers}, {Chaplin}, {Chen}, {Chiappini},
  {Chojnowski}, {Chuang}, {Clerc}, {Comparat}, {Covey}, {Croft}, {Cuesta},
  {Cunha}, {da Costa}, {Da Rio}, {Davenport}, {Dawson}, {De Lee}, {Delubac},
  {Deshpande}, {Dhital}, {Dutra-Ferreira}, {Dwelly}, {Ealet}, {Ebelke},
  {Edmondson}, {Eisenstein}, {Ellsworth}, {Elsworth}, {Epstein}, {Eracleous},
  {Escoffier}, {Esposito}, {Evans}, {Fan}, {Fern{\'a}ndez-Alvar}, {Feuillet},
  {Filiz Ak}, {Finley}, {Finoguenov}, {Flaherty}, {Fleming}, {Font-Ribera},
  {Foster}, {Frinchaboy}, {Galbraith-Frew}, {Garc{\'\i}a},
  {Garc{\'\i}a-Hern{\'a}ndez}, {Garc{\'\i}a P{\'e}rez}, {Gaulme}, {Ge},
  {G{\'e}nova-Santos}, {Georgakakis}, {Ghezzi}, {Gillespie}, {Girardi},
  {Goddard}, {Gontcho}, {Gonz{\'a}lez Hern{\'a}ndez}, {Grebel}, {Green},
  {Grieb}, {Grieves}, {Gunn}, {Guo}, {Harding}, {Hasselquist}, {Hawley},
  {Hayden}, {Hearty}, {Hekker}, {Ho}, {Hogg}, {Holley-Bockelmann}, {Holtzman},
  {Honscheid}, {Huber}, {Huehnerhoff}, {Ivans}, {Jiang}, {Johnson},
  {Kinemuchi}, {Kirkby}, {Kitaura}, {Klaene}, {Knapp}, {Kneib}, {Koenig},
  {Lam}, {Lan}, {Lang}, {Laurent}, {Le Goff}, {Leauthaud}, {Lee}, {Lee},
  {Licquia}, {Liu}, {Long}, {L{\'o}pez-Corredoira}, {Lorenzo-Oliveira},
  {Lucatello}, {Lundgren}, {Lupton}, {Mack}, {Mahadevan}, {Maia}, {Majewski},
  {Malanushenko}, {Malanushenko}, {Manchado}, {Manera}, {Mao}, {Maraston},
  {Marchwinski}, {Margala}, {Martell}, {Martig}, {Masters}, {Mathur},
  {McBride}, {McGehee}, {McGreer}, {McMahon}, {M{\'e}nard}, {Menzel},
  {Merloni}, {M{\'e}sz{\'a}ros}, {Miller}, {Miralda-Escud{\'e}}, {Miyatake},
  {Montero-Dorta}, {More}, {Morganson}, {Morice-Atkinson}, {Morrison},
  {Mosser}, {Muna}, {Myers}, {Nand ra}, {Newman}, {Neyrinck}, {Nguyen},
  {Nichol}, {Nidever}, {Noterdaeme}, {Nuza}, {O'Connell}, {O'Connell},
  {O'Connell}, {Ogando}, {Olmstead}, {Oravetz}, {Oravetz}, {Osumi}, {Owen},
  {Padgett}, {Padmanabhan}, {Paegert}, {Palanque-Delabrouille}, {Pan},
  {Parejko}, {P{\^a}ris}, {Park}, {Pattarakijwanich}, {Pellejero-Ibanez},
  {Pepper}, {Percival}, {P{\'e}rez-Fournon}, {Ṕrez-Ra`fols}, {Petitjean},
  {Pieri}, {Pinsonneault}, {Porto de Mello}, {Prada}, {Prakash},
  {Price-Whelan}, {Protopapas}, {Raddick}, {Rahman}, {Reid}, {Rich}, {Rix},
  {Robin}, {Rockosi}, {Rodrigues}, {Rodr{\'\i}guez-Torres}, {Roe}, {Ross},
  {Ross}, {Rossi}, {Ruan}, {Rubi{\~n}o-Mart{\'\i}n}, {Rykoff},
  {Salazar-Albornoz}, {Salvato}, {Samushia}, {S{\'a}nchez}, {Santiago},
  {Sayres}, {Schiavon}, {Schlegel}, {Schmidt}, {Schneider}, {Schultheis},
  {Schwope}, {Sc{\'o}ccola}, {Scott}, {Sellgren}, {Seo}, {Serenelli}, {Shane},
  {Shen}, {Shetrone}, {Shu}, {Silva Aguirre}, {Sivarani}, {Skrutskie},
  {Slosar}, {Smith}, {Sobreira}, {Souto}, {Stassun}, {Steinmetz}, {Stello},
  {Strauss}, {Streblyanska}, {Suzuki}, {Swanson}, {Tan}, {Tayar}, {Terrien},
  {Thakar}, {Thomas}, {Thomas}, {Thompson}, {Tinker}, {Tojeiro}, {Troup},
  {Vargas-Maga{\~n}a}, {Vazquez}, {Verde}, {Viel}, {Vogt}, {Wake}, {Wang},
  {Weaver}, {Weinberg}, {Weiner}, {White}, {Wilson}, {Wisniewski},
  {Wood-Vasey}, {Ye`che}, {York}, {Zakamska}, {Zamora}, {Zasowski}, {Zehavi},
  {Zhao}, {Zheng}, {Zhou}, {Zhou}, {Zou}, \& {Zhu}}]{Alam2015}
{Alam}, S., {Albareti}, F.~D., {Allende Prieto}, C., {et~al.} 2015, \apjs, 219,
  12

\bibitem[{{Barker} \& {Ogilvie}(2009)}]{BarkerOgilvie2009}
{Barker}, A.~J. \& {Ogilvie}, G.~I. 2009, \mnras, 395, 2268

\bibitem[{{Bianchi} {et~al.}(2011){Bianchi}, {Herald}, {Efremova}, {Girardi},
  {Zabot}, {Marigo}, {Conti}, \& {Shiao}}]{Bianchi2011}
{Bianchi}, L., {Herald}, J., {Efremova}, B., {et~al.} 2011, \apss, 335, 161

\bibitem[{{Boss}(1996)}]{Boss1996}
{Boss}, A.~P. 1996, in Lunar and Planetary Science Conference, Vol.~27, Lunar
  and Planetary Science Conference, 139

\bibitem[{{Boss}(1997)}]{Boss1997}
{Boss}, A.~P. 1997, Science, 276, 1836

\bibitem[{{Boss}(2002)}]{Boss2002}
{Boss}, A.~P. 2002, \apjl, 567, L149

\bibitem[{{Bouma} {et~al.}(2019){Bouma}, {Hartman}, {Bhatti}, {Winn}, \&
  {Bakos}}]{Bouma2019}
{Bouma}, L.~G., {Hartman}, J.~D., {Bhatti}, W., {Winn}, J.~N., \& {Bakos},
  G.~{\'A}. 2019, \apjs, 245, 13

\bibitem[{{Bryant} {et~al.}(2020){Bryant}, {Bayliss}, {Nielsen}, {Veras},
  {Acton}, {Anderson}, {Armstrong}, {Bouchy}, {Briegal}, {Burleigh}, {Cabrera},
  {Casewell}, {Chaushev}, {Cooke}, {Csizmadia}, {Eigmuller}, {Erikson}, {Gill},
  {Gillen}, {Goad}, {Grieves}, {Gunther}, {Henderson}, {Hogan}, {Jenkins},
  {Lendl}, {McCormac}, {Moyano}, {Queloz}, {Rauer}, {Raynard}, {Smith},
  {Tilbrook}, {Udry}, {Vines}, {Watson}, {West}, \& {Wheatley}}]{Bryant2020}
{Bryant}, E.~M., {Bayliss}, D., {Nielsen}, L.~D., {et~al.} 2020, arXiv
  e-prints, arXiv:2009.10620

\bibitem[{{Cai} {et~al.}(2006){Cai}, {Durisen}, {Michael}, {Boley},
  {Mej{\'\i}a}, {Pickett}, \& {D'Alessio}}]{Cai2006}
{Cai}, K., {Durisen}, R.~H., {Michael}, S., {et~al.} 2006, \apjl, 636, L149

\bibitem[{{Carmichael} {et~al.}(2020{\natexlab{a}}){Carmichael}, {Quinn},
  {Mustill}, {Huang}, {Zhou}, {Persson}, {Nielsen}, {Collins}, {Ziegler},
  {Collins}, {Rodriguez}, {Shporer}, {Brahm}, {Mann}, {Bouchy}, {Fridlund},
  {Stassun}, {Hellier}, {Seidel}, {Stalport}, {Udry}, {Pepe}, {Ireland},
  {{\v{Z}}erjal}, {Brice{\~n}o}, {Law}, {Jord{\'a}n}, {Espinoza}, {Henning},
  {Sarkis}, \& {Latham}}]{Carmichael2020a}
{Carmichael}, T.~W., {Quinn}, S.~N., {Mustill}, A.~J., {et~al.}
  2020{\natexlab{a}}, \aj, 160, 53

\bibitem[{{Carmichael} {et~al.}(2020{\natexlab{b}}){Carmichael}, {Quinn},
  {Zhou}, {Grieves}, {Bouchy}, {Collins}, {Kielkopf}, {Schwarz}, {Vanderburg},
  {Irwin}, {Charbonneau}, {Ziegler}, {Briceno}, {Law}, {Mann}, {Huang},
  {Shporer}, {Rodriguez}, {Stassun}, \& {Latham}}]{Carmichael2020b}
{Carmichael}, T.~W., {Quinn}, S.~N., {Zhou}, G., {et~al.} 2020{\natexlab{b}},
  arXiv e-prints, arXiv:2009.13515

\bibitem[{{Choi} {et~al.}(2016){Choi}, {Dotter}, {Conroy}, {Cantiello},
  {Paxton}, \& {Johnson}}]{Choi2016}
{Choi}, J., {Dotter}, A., {Conroy}, C., {et~al.} 2016, \apj, 823, 102

\bibitem[{{Claret} \& {Bloemen}(2011)}]{ClaretBloemen2011}
{Claret}, A. \& {Bloemen}, S. 2011, \aap, 529, A75

\bibitem[{{Cubillos} {et~al.}(2017){Cubillos}, {Harrington}, {Loredo}, {Lust},
  {Blecic}, \& {Stemm}}]{Cubillos2017}
{Cubillos}, P., {Harrington}, J., {Loredo}, T.~J., {et~al.} 2017, \aj, 153, 3

\bibitem[{{D'Angelo} {et~al.}(2003){D'Angelo}, {Kley}, \&
  {Henning}}]{DAngelo2003}
{D'Angelo}, G., {Kley}, W., \& {Henning}, T. 2003, \apj, 586, 540

\bibitem[{{Dawson} \& {Johnson}(2018)}]{DawsonJohnson2018}
{Dawson}, R.~I. \& {Johnson}, J.~A. 2018, \araa, 56, 175

\bibitem[{{Dotter}(2016)}]{Dotter2016}
{Dotter}, A. 2016, \apjs, 222, 8

\bibitem[{{Eastman}(2017)}]{Eastman2017}
{Eastman}, J. 2017, {EXOFASTv2: Generalized publication-quality exoplanet
  modeling code}

\bibitem[{{Eastman} {et~al.}(2013){Eastman}, {Gaudi}, \& {Agol}}]{Eastman2013}
{Eastman}, J., {Gaudi}, B.~S., \& {Agol}, E. 2013, \pasp, 125, 83

\bibitem[{{Eastman} {et~al.}(2019){Eastman}, {Rodriguez}, {Agol}, {Stassun},
  {Beatty}, {Vanderburg}, {Gaudi}, {Collins}, \& {Luger}}]{Eastman2019}
{Eastman}, J.~D., {Rodriguez}, J.~E., {Agol}, E., {et~al.} 2019, arXiv
  e-prints, arXiv:1907.09480

\bibitem[{{Ford}(2006)}]{Ford2006}
{Ford}, E.~B. 2006, \apj, 642, 505

\bibitem[{{Gelman} {et~al.}(2003){Gelman}, {Carlin}, {Stern}, \&
  {Rubin}}]{Gelman2003}
{Gelman}, A., {Carlin}, J.~B., {Stern}, H.~S., \& {Rubin}, D.~B. 2003, Bayesian
  Data Analysis, 2nd edn. (London: Chapman \& Hall)

\bibitem[{{Gelman} \& {Rubin}(1992)}]{GelmanRubin1992}
{Gelman}, A. \& {Rubin}, D.~B. 1992, Statistical Science, 7, 457

\bibitem[{{Gill} {et~al.}(2019){Gill}, {Maxted}, {Evans}, {Evans},
  {Southworth}, {Smalley}, {Gary}, {Anderson}, {Bouchy}, {Cameron}, {Dominik},
  {Faedi}, {Gillon}, {Gomez Maqueo Chew}, {Hebb}, {Hellier}, {J{\o}rgensen},
  {Longa-Pe{\~n}a}, {Martin}, {McCormac}, {Pepe}, {Pollaco}, {Queloz},
  {S{\'e}gransan}, {Snodgrass}, {Turner}, {Triaud}, {Udry}, \&
  {West}}]{Gill2019}
{Gill}, S., {Maxted}, P.~F.~L., {Evans}, J.~A., {et~al.} 2019, \aap, 626, A119

\bibitem[{{Gill} {et~al.}(2018){Gill}, {Maxted}, \& {Smalley}}]{Gill2018}
{Gill}, S., {Maxted}, P.~F.~L., \& {Smalley}, B. 2018, \aap, 612, A111

\bibitem[{{Gill} {et~al.}(2020){Gill}, {Wheatley}, {Cooke}, {Jord{\'a}n},
  {Nielsen}, {Bayliss}, {Anderson}, {Vines}, {Lendl}, {Acton}, {Armstrong},
  {Bouchy}, {Brahm}, {Bryant}, {Burleigh}, {Casewell}, {Eigm{\"u}ller},
  {Espinoza}, {Gillen}, {Goad}, {Grieves}, {G{\"u}nther}, {Henning}, {Hobson},
  {Hogan}, {Jenkins}, {McCormac}, {Moyano}, {Osborn}, {Pollacco}, {Queloz},
  {Rauer}, {Raynard}, {Rojas}, {Sarkis}, {Smith}, {Pinto}, {Tilbrook}, {Udry},
  {Watson}, \& {West}}]{Gill2020}
{Gill}, S., {Wheatley}, P.~J., {Cooke}, B.~F., {et~al.} 2020, \apjl, 898, L11

\bibitem[{{Goda} \& {Matsuo}(2019)}]{GodaMatsuo2019}
{Goda}, S. \& {Matsuo}, T. 2019, \apj, 876, 23

\bibitem[{{Grether} \& {Lineweaver}(2006)}]{Grether2006}
{Grether}, D. \& {Lineweaver}, C.~H. 2006, \apj, 640, 1051

\bibitem[{{Grunblatt} {et~al.}(2019){Grunblatt}, {Huber}, {Gaidos}, {Hon},
  {Zinn}, \& {Stello}}]{Grunblatt2019}
{Grunblatt}, S.~K., {Huber}, D., {Gaidos}, E., {et~al.} 2019, \aj, 158, 227

\bibitem[{{Grunblatt} {et~al.}(2018){Grunblatt}, {Huber}, {Gaidos}, {Lopez},
  {Barclay}, {Chontos}, {Sinukoff}, {Van Eylen}, {Howard}, \&
  {Isaacson}}]{Grunblatt2018}
{Grunblatt}, S.~K., {Huber}, D., {Gaidos}, E., {et~al.} 2018, \apjl, 861, L5

\bibitem[{{Hansen}(2010)}]{Hansen2010}
{Hansen}, B. M.~S. 2010, \apj, 723, 285

\bibitem[{{Hara} {et~al.}(2019){Hara}, {Bou{\'e}}, {Laskar}, {Delisle}, \&
  {Unger}}]{Hara2019}
{Hara}, N.~C., {Bou{\'e}}, G., {Laskar}, J., {Delisle}, J.~B., \& {Unger}, N.
  2019, \mnras, 489, 738

\bibitem[{{Henden} {et~al.}(2016){Henden}, {Templeton}, {Terrell}, {Smith},
  {Levine}, \& {Welch}}]{Henden2016}
{Henden}, A.~A., {Templeton}, M., {Terrell}, D., {et~al.} 2016, VizieR Online
  Data Catalog, II/336

\bibitem[{{Hut}(1981)}]{Hut1981}
{Hut}, P. 1981, \aap, 99, 126

\bibitem[{{Ikwut-Ukwa} {et~al.}(2020){Ikwut-Ukwa}, {Rodriguez}, {Bieryla},
  {Vanderburg}, {Mocnik}, {Kane}, {Quinn}, {Col{\'o}n}, {Zhou}, {Eastman},
  {Huang}, {Latham}, {Dotson}, {Jenkins}, {Ricker}, {Seager}, {Vanderspek},
  {Winn}, {Barclay}, {Barentsen}, {Berta-Thompson}, {Charbonneau}, {Dragomir},
  {Daylan}, {G{\"u}nther}, {Hedges}, {Henze}, {McDermott}, {Schlieder},
  {Quintana}, {Smith}, {Twicken}, \& {Yahalomi}}]{Ikwut-Ukwa2020}
{Ikwut-Ukwa}, M., {Rodriguez}, J.~E., {Bieryla}, A., {et~al.} 2020, \aj, 160,
  209

\bibitem[{{Kov{\'a}cs} {et~al.}(2002){Kov{\'a}cs}, {Zucker}, \&
  {Mazeh}}]{Kovacs2002}
{Kov{\'a}cs}, G., {Zucker}, S., \& {Mazeh}, T. 2002, \aap, 391, 369

\bibitem[{{Lee} \& {Chiang}(2016)}]{LeeChiang2016}
{Lee}, E.~J. \& {Chiang}, E. 2016, \apj, 817, 90

\bibitem[{{Lee} {et~al.}(2014){Lee}, {Chiang}, \& {Ormel}}]{Lee2014}
{Lee}, E.~J., {Chiang}, E., \& {Ormel}, C.~W. 2014, \apj, 797, 95

\bibitem[{{Lillo-Box} {et~al.}(2016){Lillo-Box}, {Barrado}, \&
  {Correia}}]{Lillo-Box2016}
{Lillo-Box}, J., {Barrado}, D., \& {Correia}, A.~C.~M. 2016, \aap, 589, A124

\bibitem[{{Lin} {et~al.}(1996){Lin}, {Bodenheimer}, \& {Richardson}}]{Lin1996}
{Lin}, D.~N.~C., {Bodenheimer}, P., \& {Richardson}, D.~C. 1996, \nat, 380, 606

\bibitem[{{Lucy} \& {Sweeney}(1971)}]{Lucy1971}
{Lucy}, L.~B. \& {Sweeney}, M.~A. 1971, \aj, 76, 544

\bibitem[{{Ma} \& {Ge}(2014)}]{MaGe2014}
{Ma}, B. \& {Ge}, J. 2014, \mnras, 439, 2781

\bibitem[{{Maldonado} {et~al.}(2019){Maldonado}, {Villaver}, {Eiroa}, \&
  {Micela}}]{Maldonado2019}
{Maldonado}, J., {Villaver}, E., {Eiroa}, C., \& {Micela}, G. 2019, \aap, 624,
  A94

\bibitem[{{Marcy} \& {Butler}(2000)}]{MarcyButler2000}
{Marcy}, G.~W. \& {Butler}, R.~P. 2000, \pasp, 112, 137

\bibitem[{{Mayor} \& {Queloz}(1995)}]{MayorQueloz1995}
{Mayor}, M. \& {Queloz}, D. 1995, \nat, 378, 355

\bibitem[{{McLaughlin}(1924)}]{McLaughlin1924}
{McLaughlin}, D.~B. 1924, \apj, 60, 22

\bibitem[{{Mireles} {et~al.}(2020){Mireles}, {Shporer}, {Grieves}, {Zhou},
  {G{\"u}nther}, {Brahm}, {Ziegler}, {Stassun}, {Huang}, {Nielsen}, {dos
  Santos}, {Udry}, {Bouchy}, {Ireland }, {Wallace}, {Sarkis}, {Henning},
  {Jordan}, {Law}, {Mann}, {Paredes}, {James}, {Jao}, {Henry}, {Butler},
  {Rodriguez}, {Yu}, {Flowers}, {Ricker}, {Latham}, {Vand erspek}, {Seager},
  {Winn}, {Jenkins}, {Furesz}, {Hesse}, {Quintana}, {Rose}, {Smith},
  {Tenenbaum}, {Vezie}, {Yahalomi}, \& {Zhan}}]{Mireles2020}
{Mireles}, I., {Shporer}, A., {Grieves}, N., {et~al.} 2020, arXiv e-prints,
  arXiv:2006.14019

\bibitem[{{P{\'a}l}(2009)}]{Pal2009}
{P{\'a}l}, A. 2009, PhD thesis, Department of Astronomy, E{\"o}tv{\"o}s
  Lor{\'a}nd University

\bibitem[{{Pecaut} \& {Mamajek}(2013)}]{PecautMamajek2013}
{Pecaut}, M.~J. \& {Mamajek}, E.~E. 2013, \apjs, 208, 9

\bibitem[{{Pepe} {et~al.}(2002){Pepe}, {Mayor}, {Rupprecht}, {Avila},
  {Ballester}, {Beckers}, {Benz}, {Bertaux}, {Bouchy}, {Buzzoni}, {Cavadore},
  {Deiries}, {Dekker}, {Delabre}, {D'Odorico}, {Eckert}, {Fischer}, {Fleury},
  {George}, {Gilliotte}, {Gojak}, {Guzman}, {Koch}, {Kohler}, {Kotzlowski},
  {Lacroix}, {Le Merrer}, {Lizon}, {Lo Curto}, {Longinotti}, {Megevand},
  {Pasquini}, {Petitpas}, {Pichard}, {Queloz}, {Reyes}, {Richaud}, {Sivan},
  {Sosnowska}, {Soto}, {Udry}, {Ureta}, {van Kesteren}, {Weber}, {Weilenmann},
  {Wicenec}, {Wieland}, {Christensen-Dalsgaard}, {Dravins}, {Hatzes},
  {K{\"u}rster}, {Paresce}, \& {Penny}}]{Pepe2002}
{Pepe}, F., {Mayor}, M., {Rupprecht}, G., {et~al.} 2002, The Messenger, 110, 9

\bibitem[{{Pepper} {et~al.}(2020){Pepper}, {Kane}, {Rodriguez}, {Hinkel},
  {Eastman}, {Daylan}, {Mocnik}, {Dalba}, {Gaudi}, {Fetherolf}, {Stassun},
  {Campante}, {Vanderburg}, {Huber}, {Bossini}, {Crossfield}, {Howell},
  {Stephens}, {Furlan}, {Ricker}, {Vanderspek}, {Latham}, {Seager}, {Winn},
  {Jenkins}, {Twicken}, {Rose}, {Smith}, {Glidden}, {Levine}, {Rinehart},
  {Collins}, {Mann}, {Burt}, {James}, {Siverd}, \& {G{\"u}nther}}]{Pepper2020}
{Pepper}, J., {Kane}, S.~R., {Rodriguez}, J.~E., {et~al.} 2020, \aj, 159, 243

\bibitem[{{Piso} {et~al.}(2015){Piso}, {Youdin}, \& {Murray-Clay}}]{Piso2015}
{Piso}, A.-M.~A., {Youdin}, A.~N., \& {Murray-Clay}, R.~A. 2015, \apj, 800, 82

\bibitem[{{Pollack} {et~al.}(1996){Pollack}, {Hubickyj}, {Bodenheimer},
  {Lissauer}, {Podolak}, \& {Greenzweig}}]{Pollack1996}
{Pollack}, J.~B., {Hubickyj}, O., {Bodenheimer}, P., {et~al.} 1996, \icarus,
  124, 62

\bibitem[{{Queloz} {et~al.}(2001){Queloz}, {Mayor}, {Udry}, {Burnet},
  {Carrier}, {Eggenberger}, {Naef}, {Santos}, {Pepe}, {Rupprecht}, {Avila},
  {Baeza}, {Benz}, {Bertaux}, {Bouchy}, {Cavadore}, {Delabre}, {Eckert},
  {Fischer}, {Fleury}, {Gilliotte}, {Goyak}, {Guzman}, {Kohler}, {Lacroix},
  {Lizon}, {Megevand}, {Sivan}, {Sosnowska}, \& {Weilenmann}}]{Queloz2001}
{Queloz}, D., {Mayor}, M., {Udry}, S., {et~al.} 2001, The Messenger, 105, 1

\bibitem[{{Rafikov}(2006)}]{Rafikov2006}
{Rafikov}, R.~R. 2006, \apj, 648, 666

\bibitem[{{Rasio} \& {Ford}(1996)}]{RasioFord1996}
{Rasio}, F.~A. \& {Ford}, E.~B. 1996, Science, 274, 954

\bibitem[{{Ricker} {et~al.}(2015){Ricker}, {Winn}, {Vanderspek}, {Latham},
  {Bakos}, {Bean}, {Berta-Thompson}, {Brown}, {Buchhave}, {Butler}, {Butler},
  {Chaplin}, {Charbonneau}, {Christensen-Dalsgaard}, {Clampin}, {Deming},
  {Doty}, {De Lee}, {Dressing}, {Dunham}, {Endl}, {Fressin}, {Ge}, {Henning},
  {Holman}, {Howard}, {Ida}, {Jenkins}, {Jernigan}, {Johnson}, {Kaltenegger},
  {Kawai}, {Kjeldsen}, {Laughlin}, {Levine}, {Lin}, {Lissauer}, {MacQueen},
  {Marcy}, {McCullough}, {Morton}, {Narita}, {Paegert}, {Palle}, {Pepe},
  {Pepper}, {Quirrenbach}, {Rinehart}, {Sasselov}, {Sato}, {Seager},
  {Sozzetti}, {Stassun}, {Sullivan}, {Szentgyorgyi}, {Torres}, {Udry}, \&
  {Villasenor}}]{Ricker2015}
{Ricker}, G.~R., {Winn}, J.~N., {Vanderspek}, R., {et~al.} 2015, Journal of
  Astronomical Telescopes, Instruments, and Systems, 1, 014003

\bibitem[{{Rossiter}(1924)}]{Rossiter1924}
{Rossiter}, R.~A. 1924, \apj, 60, 15

\bibitem[{{Santos} {et~al.}(2017){Santos}, {Adibekyan}, {Figueira},
  {Andreasen}, {Barros}, {Delgado-Mena}, {Demangeon}, {Faria}, {Oshagh},
  {Sousa}, {Viana}, \& {Ferreira}}]{Santos2017}
{Santos}, N.~C., {Adibekyan}, V., {Figueira}, P., {et~al.} 2017, \aap, 603, A30

\bibitem[{{Schlafly} \& {Finkbeiner}(2011)}]{Schlafly2011}
{Schlafly}, E.~F. \& {Finkbeiner}, D.~P. 2011, \apj, 737, 103

\bibitem[{{Schlaufman}(2018)}]{Schlaufman2018}
{Schlaufman}, K.~C. 2018, \apj, 853, 37

\bibitem[{{Sestovic} {et~al.}(2018){Sestovic}, {Demory}, \&
  {Queloz}}]{Sestovic2018}
{Sestovic}, M., {Demory}, B.-O., \& {Queloz}, D. 2018, \aap, 616, A76

\bibitem[{{Skrutskie} {et~al.}(2006){Skrutskie}, {Cutri}, {Stiening},
  {Weinberg}, {Schneider}, {Carpenter}, {Beichman}, {Capps}, {Chester},
  {Elias}, {Huchra}, {Liebert}, {Lonsdale}, {Monet}, {Price}, {Seitzer},
  {Jarrett}, {Kirkpatrick}, {Gizis}, {Howard}, {Evans}, {Fowler}, {Fullmer},
  {Hurt}, {Light}, {Kopan}, {Marsh}, {McCallon}, {Tam}, {Van Dyk}, \&
  {Wheelock}}]{Skrutskie2006}
{Skrutskie}, M.~F., {Cutri}, R.~M., {Stiening}, R., {et~al.} 2006, \aj, 131,
  1163

\bibitem[{{Stassun} {et~al.}(2019){Stassun}, {Oelkers}, {Paegert}, {Torres},
  {Pepper}, {De Lee}, {Collins}, {Latham}, {Muirhead}, {Chittidi},
  {Rojas-Ayala}, {Fleming}, {Rose}, {Tenenbaum}, {Ting}, {Kane}, {Barclay},
  {Bean}, {Brassuer}, {Charbonneau}, {Ge}, {Lissauer}, {Mann}, {McLean},
  {Mullally}, {Narita}, {Plavchan}, {Ricker}, {Sasselov}, {Seager}, {Sharma},
  {Shiao}, {Sozzetti}, {Stello}, {Vanderspek}, {Wallace}, \&
  {Winn}}]{Stassun2019}
{Stassun}, K.~G., {Oelkers}, R.~J., {Paegert}, M., {et~al.} 2019, \aj, 158, 138

\bibitem[{{Stassun} \& {Torres}(2016)}]{StassunTorres2016}
{Stassun}, K.~G. \& {Torres}, G. 2016, \aj, 152, 180

\bibitem[{{Tamuz} {et~al.}(2005){Tamuz}, {Mazeh}, \& {Zucker}}]{Tamuz2005}
{Tamuz}, O., {Mazeh}, T., \& {Zucker}, S. 2005, \mnras, 356, 1466

\bibitem[{{Triaud} {et~al.}(2010){Triaud}, {Collier Cameron}, {Queloz},
  {Anderson}, {Gillon}, {Hebb}, {Hellier}, {Loeillet}, {Maxted}, {Mayor},
  {Pepe}, {Pollacco}, {S{\'e}gransan}, {Smalley}, {Udry}, {West}, \&
  {Wheatley}}]{Triaud2010}
{Triaud}, A.~H.~M.~J., {Collier Cameron}, A., {Queloz}, D., {et~al.} 2010,
  \aap, 524, A25

\bibitem[{{Wheatley} {et~al.}(2018){Wheatley}, {West}, {Goad}, {Jenkins},
  {Pollacco}, {Queloz}, {Rauer}, {Udry}, {Watson}, {Chazelas}, {Eigm{\"u}ller},
  {Lambert}, {Genolet}, {McCormac}, {Walker}, {Armstrong}, {Bayliss}, {Bento},
  {Bouchy}, {Burleigh}, {Cabrera}, {Casewell}, {Chaushev}, {Chote},
  {Csizmadia}, {Erikson}, {Faedi}, {Foxell}, {G{\"a}nsicke}, {Gillen},
  {Grange}, {G{\"u}nther}, {Hodgkin}, {Jackman}, {Jord{\'a}n}, {Louden},
  {Metrailler}, {Moyano}, {Nielsen}, {Osborn}, {Poppenhaeger}, {Raddi},
  {Raynard}, {Smith}, {Soto}, \& {Titz-Weider}}]{Wheatley2018}
{Wheatley}, P.~J., {West}, R.~G., {Goad}, M.~R., {et~al.} 2018, \mnras, 475,
  4476

\bibitem[{{Winn} {et~al.}(2005){Winn}, {Noyes}, {Holman}, {Charbonneau},
  {Ohta}, {Taruya}, {Suto}, {Narita}, {Turner}, {Johnson}, {Marcy}, {Butler},
  \& {Vogt}}]{Winn2005}
{Winn}, J.~N., {Noyes}, R.~W., {Holman}, M.~J., {et~al.} 2005, \apj, 631, 1215

\bibitem[{{Wright} {et~al.}(2010){Wright}, {Eisenhardt}, {Mainzer}, {Ressler},
  {Cutri}, {Jarrett}, {Kirkpatrick}, {Padgett}, {McMillan}, {Skrutskie},
  {Stanford}, {Cohen}, {Walker}, {Mather}, {Leisawitz}, {Gautier}, {McLean},
  {Benford}, {Lonsdale}, {Blain}, {Mendez}, {Irace}, {Duval}, {Liu}, {Royer},
  {Heinrichsen}, {Howard}, {Shannon}, {Kendall}, {Walsh}, {Larsen}, {Cardon},
  {Schick}, {Schwalm}, {Abid}, {Fabinsky}, {Naes}, \& {Tsai}}]{Wright2010}
{Wright}, E.~L., {Eisenhardt}, P. R.~M., {Mainzer}, A.~K., {et~al.} 2010, \aj,
  140, 1868

\bibitem[{{Yee} {et~al.}(2017){Yee}, {Petigura}, \& {von Braun}}]{Yee2017}
{Yee}, S.~W., {Petigura}, E.~A., \& {von Braun}, K. 2017, \apj, 836, 77

\bibitem[{{Zhou} {et~al.}(2019){Zhou}, {Huang}, {Bakos}, {Hartman}, {Latham},
  {Quinn}, {Collins}, {Winn}, {Wong}, {Kov{\'a}cs}, {Csubry}, {Bhatti},
  {Penev}, {Bieryla}, {Esquerdo}, {Berlind}, {Calkins}, {de Val-Borro},
  {Noyes}, {L{\'a}z{\'a}r}, {Papp}, {S{\'a}ri}, {Kov{\'a}cs}, {Buchhave},
  {Szklenar}, {B{\'e}ky}, {Johnson}, {Cochran}, {Kniazev}, {Stassun}, {Fulton},
  {Shporer}, {Espinoza}, {Bayliss}, {Everett}, {Howell}, {Hellier}, {Anderson},
  {Collier Cameron}, {West}, {Brown}, {Schanche}, {Barkaoui}, {Pozuelos},
  {Gillon}, {Jehin}, {Benkhaldoun}, {Daassou}, {Ricker}, {Vanderspek},
  {Seager}, {Jenkins}, {Lissauer}, {Armstrong}, {Collins}, {Gan}, {Hart},
  {Horne}, {Kielkopf}, {Nielsen}, {Nishiumi}, {Narita}, {Palle}, {Relles},
  {Sefako}, {Tan}, {Davies}, {Goeke}, {Guerrero}, {Haworth}, \&
  {Villanueva}}]{Zhou2019}
{Zhou}, G., {Huang}, C.~X., {Bakos}, G.~{\'A}., {et~al.} 2019, \aj, 158, 141

\end{thebibliography}

\end{document}